\documentclass[%
aps,
pre,%
amsmath,amssymb,
reprint,%
]{revtex4-1}
\usepackage{color}
\usepackage[dvipsnames]{xcolor}
\usepackage{graphicx}
\usepackage{dcolumn}
\usepackage{bm}
\usepackage[bookmarks=false,hyperfootnotes=false]{hyperref}
\hypersetup{
	colorlinks=true,
	linkcolor=blue,
	anchorcolor=black,
	citecolor=blue,
	urlcolor=blue,
}
\usepackage{multirow}
\usepackage{amssymb,amsfonts,amsmath}
\usepackage{mathtools,mathrsfs}
\usepackage{placeins} 
\usepackage{soul}
\usepackage{enumerate}   

\newcommand{\dt}{\partial_{t} }

\newcommand{\dey}{\partial_{y} }

\newcommand{\ddey}{\partial^{2}_{y} }
\newcommand{\dddey}{\partial^{3}_{y} }
\newcommand{\ddddey}{\partial^{4}_{y} }
\newcommand{\dtot}[2]{\frac{\mathrm{d} #1}{\mathrm{d} #2}}
\definecolor{amber}{rgb}{1.0, 0.75, 0.0}
\newcommand{\ud}{\mathop{}\!\mathrm{d}}  
\begin{document}
	
	
	\title{Linear waves in sheared flows. Lower bound of the vorticity growth and  propagation discontinuities in the parameters space} 
	
	
	
	\author{Federico Fraternale$^1$, Loris Domenicale$^2$, Gigliola Staffilani$^3$ , Daniela Tordella$^1$}
	\thanks{Email address for correspondence: daniela.tordella@polito.it}
	\affiliation{$^1$ Department of Applied Science and Technology, Politecnico di Torino, Torino, Italy 10129}
	\affiliation{$^2$ Department of Mathematical Sciences``G. L. Lagrange", Politecnico di Torino, Torino, Italy 10129}\altaffiliation{currently at University of Southampton, Faculty of Engineering \& Environment, Southampton SO16 7QF, England}
	\affiliation{$^3$ Department of Mathematics, Massachusetts Institute of Technology,  Cambridge, MA 02139-4307, USA}
	
	\date{\today}

\begin{abstract}
This study provides sufficient conditions for the temporal monotonic decay of enstrophy for two-dimensional perturbations traveling in the incompressible, viscous, plane Poiseuille and Couette flows. Extension of J. L. Synge's procedure (1938) to the initial-value problem allowed us to find the region of the wavenumber-Reynolds number map where the enstrophy of any initial disturbance cannot grow. This region is wider than the kinetic energy's one. We also show that the parameters space is split in two regions with clearly distinct propagation and dispersion properties.
	


\end{abstract}

\maketitle


\section{Introduction}
One reason for the limited use of the enstrophy quantity \cite{frisch1995,tsinober2001book,abramov2003,yeung2015,schumacher2015} in hydrodynamic stability theory of wall flows is certainly the lack of knowledge of physical  boundary conditions on the vorticity \cite{synge1935}. Oppositely, for the velocity field the wall boundary conditions (no-slip) have been known for more than a century, at least for the wide class of wall flows under the continuum hypothesis. Notwithstanding this, our work is focused on the enstrophy of  traveling perturbation waves in wall flows.  One objective of this study is to highlight the role of the enstrophy as well as its interrelationship with the more commonly considered kinetic energy. In particular,  we are interested in the conditions  for  transient growth of the perturbation's enstrophy in the wavenumber - Reynolds number parameters space.

We consider the two-dimensional plane Poiseuille and Couette flows, which are emblematic problems of the hydrodynamic stability theory. \textcolor{black}{The flow velocity field can be decomposed in a basic laminar state and a fluctuation about it, that is $\mathbf{u}(t,\mathbf{x})=\mathbf{U}(\mathbf{x})+ \mathbf{\tilde u}(t,\mathbf{x})$}. Departures from the basic state may appear  in the subcritical range below the critical Reynolds number $Re_\mathrm{c}$ that is the lower limit  for \textit{unconditional instability} (notice that throughout this discussion we adopt the terminology used by Manneville \cite{manneville2016}). Furthermore, the kinetic energy method \cite{joseph1976} generates a lower bound  to the \textit{unconditional} (or \textit{global}) \textit{stability threshold} represented by the value $Re_\mathrm{g}$ \cite{chapman2002,manneville2015a,manneville2016}. The condition defining $Re_\mathrm{g}$  stands on the ultimate perturbations decay of both kinetic energy and enstrophy, whatever the initial disturbance amplitude and the transient growth experienced in the intermediate term. Values for the 3D case, collected from experiments in the literature, are around 325 for the plane Couette flow and 840 for the plane Poiseuille flow.  \cite{orszag1980,carlson1982,lundbladh1991,daviaud1992,ttrd1993,waleffe1995a,waleffe1995b,bottin1998,bottin1998b,duguet2010,manneville2012,tsukahara2014,manneville2015a}. A lower bound \textcolor{black}{for $Re_\mathrm{g}$}, named $Re_\mathrm{E}$, specifies instead the value below which the kinetic energy of any perturbation inside the basic flow decreases monotonically to zero. 

In this paper, \textit{for any possible initial condition} we obtain the limiting curve in the stability map wavenumber ($\alpha$) - Reynolds  for the monotonic decay of the integral enstrophy of two-dimensional perturbations. We show that this bound is less restrictive than the limiting curve for the kinetic energy decay. \textcolor{black}{In this regard, it should be recalled that, {\bf in two dimensions}, due to the absence of the vortex stretching-tilting mechanism, enstrophy is an inviscid invariant as it is in general the kinetic energy for any dimension.  Therefore, in two dimensions, the rate of change of the total enstrophy behaves  in a similar way to the total  kinetic energy $E$ of a disturbance  as described by the Reynolds-Orr equation  (see \cite{rh1993},  \cite{henningson1993}, \cite{henningson1996},  \cite{sh2001}).}

The monotonic decay region for the enstrophy in the stability map is obtained by extending the non-modal approach to a procedure proposed in 1938 by J. L. Synge in a proceeding paper of the London Mathematical Society \cite{synge1938b} that  has not been further exploited. Synge's procedure was aimed at finding analytical conditions satisfied by both the vorticity and the stream function in the two-dimensional plane Poiseuille flow \cite{synge1935,synge1938a,synge1938b}.  The  procedure is based on the deduction of the cross derivative of the flow vorticity by using the Orr-Sommerfeld equation,  which is then coupled to an optimization process acting directly of the vorticity integral. 

\textcolor{black}{The impact of this result is that the lower bound for perturbation transient growth is improved if we consider the problem in terms of enstrophy instead of kinetic energy. These results are contextualized within the structure of the stability map ($\alpha,Re$), which also contains information of dispersion properties of the least stable perturbation}. 

\textcolor{black}{The paper is organized in the following way. In Sec. II, both the relationship between enstrophy and kinetic energy and the procedure to obtain the lower bound for the transient growth of perturbation enstrophy are described (see also the Appendix). Results concerning the structure of stability map and related wave dispersion properties are  discussed in Sec. III.  Conclusion remarks follow in Sec. IV. The Appendix contains the analytical calculations leading to the lower bound. In the Supplemental Material, the reader can find the map describing the timing of maximal growth of kinetic energy and enstrophy (S1), information on the temporal evolution of enstrophy-rate optimal streamfunctions (S2), Mathematica scripts (S3). 
For the sake of brevity, in the following, the plane Poiseuille and plane Couette flows may be referred to as PPF and PCF, respectively.}

\section{Relationship between enstrophy and energy of small internal waves in parallel flows. The problem of transient enstrophy growth.}

\begin{figure}
	\centering
	\includegraphics[width=.99\columnwidth]{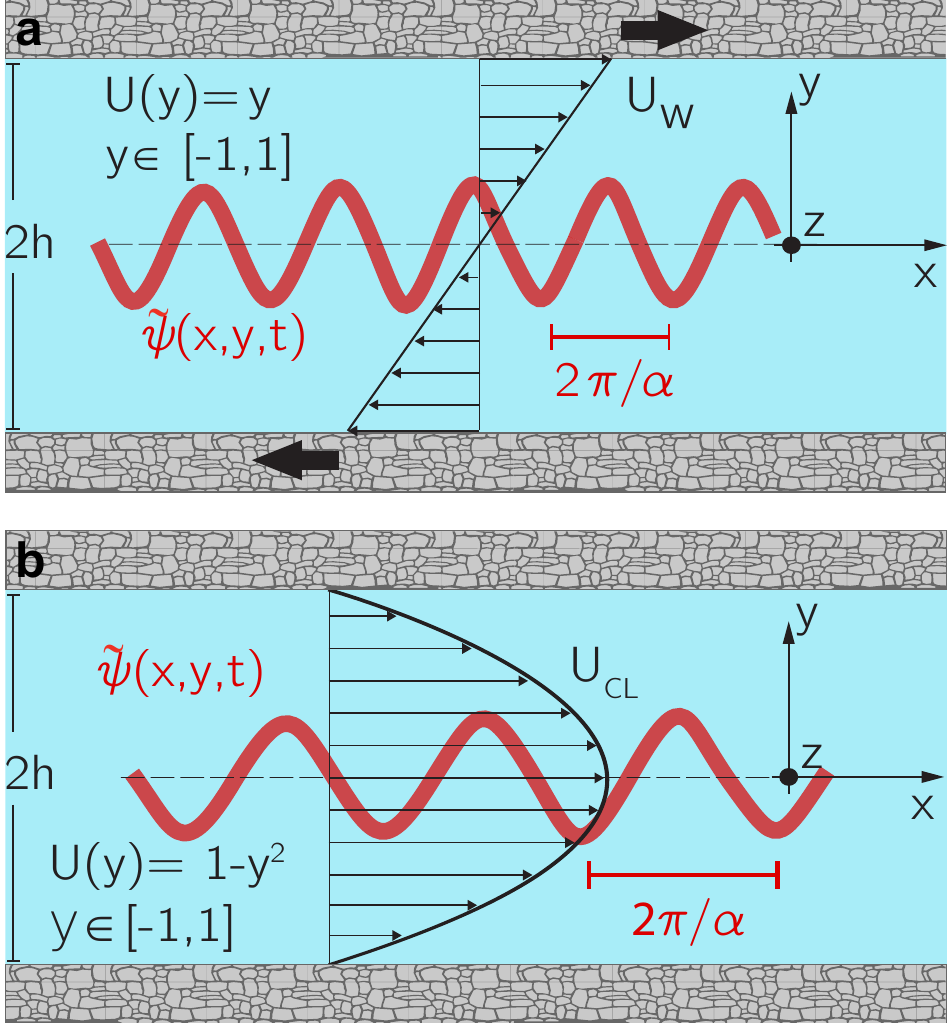}
	\vskip-10pt\caption{\textbf{Sketch of basic flows, reference systems and reference quantities}. \textbf{(a)} The plane Couette flow (PCF), a flow driven by the reciprocal sliding of two solid walls. The reference length is the channel half-height $h$, while the reference velocity is    $U_\mathrm{w}$, that is half the relative speed between the walls. The flow control parameter is the Reynolds number $Re=U_\mathrm{w}h/\nu$, where $\nu$ is the kinematic viscosity. \textbf{(b)} The plane Poiseuille flow (PPF) is a flow between two fixed walls, driven by the pressure gradient along the channel axis. Here, the reference velocity is the centerline velocity $U_\mathrm{CL}$, thus  $Re=U_\mathrm{CL}h/\nu$. The Cartesian reference system  is located at the channel centerline. The red oscillation represents a perturbation with wavenumber $\alpha$. }\vskip-10pt
	\label{fig:base_flow}
\end{figure}

\textcolor{black}{The plane Couette flow is a parallel viscous taking place in the gap between two plates moving tangentially one respect to the other. In the following, in regard to normalization issues, the reference length is the channel half-width $h$ while the reference velocity is the walls speed half-difference, $U_{w}$. The flow control parameter is the Reynolds number $Re=U_\mathrm{w}h/\nu$, where $\nu$ is the kinematic viscosity.  The plane Poiseuille flow (PPF) is a flow driven by a pressure gradient between two fixed walls. Here, the reference velocity is the centerline velocity $U_\mathrm{CL}$, thus  $Re=U_\mathrm{CL}h/\nu$. The Cartesian reference system is located at the channel centerline. In particular, we consider the two-dimensional configuration. The domain is bounded in the cross-flow direction ($-1<y<1$) and it is unbounded in the streamwise direction $x$. The basic flow expression is $\mathbf{U}(\mathbf{x})=U(y)\mathbf{e_\mathrm{x}}$, where $U(y)=y$ for the plane Couette and $U(y)=1-y^2$ for the plane Poiseuille flow, see Fig. \ref{fig:base_flow}.}

Let us introduce the integral enstrophy for a  two-dimensional perturbation in a parallel-flow field:
\vspace{-5pt}
\begin{equation}
\Omega=\frac{1}{2\mathcal{|V|}}\int_{\mathcal{V}} \tilde\omega^2 \ud x\ud y,
\end{equation}
were   $\mathcal{V}$ is an arbitrary spatial domain and $\mathcal{|V|}$ its volume.
\begin{equation}\label{eq:vort_def_physical}
\tilde{ \omega}=\partial_x \tilde{v} - \partial_y \tilde{u}
\end{equation}
is the vorticity of the perturbation velocity field of components $\tilde u$ (longitudinal) and $\tilde v$ (wall-normal).  \\
Since we are interested in the evolution equation of the integral enstrophy $\Omega$, it is convenient to consider the non-dimensional, linearized, viscous vorticity equation for small  disturbances
\begin{equation}
\partial_t \tilde{\omega}-\tilde{v} U''+U\partial_x \tilde{\omega}=Re^{-1}\nabla^2\tilde{\omega},
\end{equation}
where the prime symbol stands for a total y-derivative. 
The integral enstrophy evolution equation is then
\begin{align}
&\frac{\ud}{\ud t} \Omega = \frac{1}{\mathcal{|V|}}\frac{\ud}{\ud t} \int_\mathcal{V}{\left( \cfrac{\tilde{\omega}^2}{2}\right)\ud x\ud y}\nonumber\\& =\frac{1}{|\mathcal{V}|}\int_\mathcal{V}\left(\tilde{v}U''\tilde{\omega}-U\tilde{\omega}\partial_x \tilde{\omega}+{Re}^{-1}\tilde{\omega}\nabla^2\tilde{\omega} \right) \ud x\ud y.
\end{align}

\textcolor{black}{Before going any further, it should be recalled why the linearized equation of motion is used to seek a lower bound on the enstrophy transient growth for perturbations of any shape and amplitude. As a matter of fact, in the two-dimensional case, the instantaneous integral-enstrophy rate $\ud Z/\ud t=\Omega^{-1}\ud \Omega/\ud t$  is independent on the perturbations amplitude. Thus, it depends on linear mechanisms only. Physically, this fact is linked to the lack of vortex stretching in 2D (a detailed discussion can be found in App. C of Tsinober's monograph,  \cite{tsinober2001book}). In contrast, in three dimensions the enstrophy is not an inviscid invariant. Here, in fact, the vortex stretching terms are responsible for the self-amplification and tilting of the vorticity. Such terms appear as cubic terms in the 3D integral-enstrophy equation for the perturbation, and make the enstrophy rate depend on the amplitude of disturbances. These terms correspond to a possible net enstrophy production and are empirically known to be positive both from laboratory and numerical experiments \cite{taylor1938,betchov1976,tsinober1998}. 
In the integral kinetic energy equation instead, both in two and three dimensions, the growth rate, $\ud G/\ud t=E^{-1}\ud E/\ud t$, is independent on the perturbation amplitude since in the Reynolds-Orr equation for finite-amplitude perturbations the cubic terms can be written in a conservative formand do not give a net contribution when integrated over a domain with homogeneous or periodic boundary conditions (see \cite{rh1993},  \cite{henningson1993}, \cite{henningson1996},  \cite{sh2001}). This explains why the limit $Re_\mathrm{E}$ given by linear analysis is actually considered a lower bound for the global instability limit $Re_\mathrm{g}$.}

\textcolor{black}{The present study does not draw conclusions on the 3D global stability, however it suggests an improvement of the lower bound for 2D global stability. }

In the following, we introduce the stream-function of the perturbation ($\tilde u=\partial_y\tilde\psi$, $\tilde v=-\partial_x\tilde\psi$)  and adopt the Fourier representation. For the generic variable $\widetilde q(x,y,t)$, we will thus consider the wave solution $\hat q(t,y;\alpha)=\tilde q(t,x,y)e^{-i\alpha x}$, 
where $i$ is the imaginary unit and $\alpha$ is the wavenumber. 

The equation for the evolution of small-amplitude wave perturbations in 2D is known as the Orr-Sommerfeld equation  (OS). In terms of the perturbation stream-function $\hat{\psi}(y,t)$, it becomes
\begin{flalign}
\label{eq:OrrSomm}
\dt (\ddey \hat\psi-\alpha^2\hat\psi) =&-i\alpha U (\ddey\hat\psi-\alpha^2\hat\psi)+i\alpha U''\hat\psi\nonumber\\
&+\frac{1}{Re}(\ddddey\hat\psi-2\alpha^2\ddey\hat\psi+\alpha^4\hat\psi).
\end{flalign}
The initial-value problem is then formulated by adding the initial condition $\hat\psi(y,t=0)=\hat\psi_0(y)$ and the no-slip boundary conditions, $\hat\psi(\pm 1,t)=\dey\hat\psi(\pm 1,t)=0$.
%
%

The wave local enstrophy can be written as
\begin{align}
&\|\hat\omega\|^2=\| i\alpha \hat{v}-\dey\hat u \|^2=\|\alpha^2 \hat{\psi} -\ddey\hat\psi\|^2\\\nonumber
&=\alpha^4 \|\hat{\psi}\|^2+\|\ddey\hat\psi\|^2-2\alpha^2 \Re(\hat{\psi})\mathfrak{R}(\ddey\hat\psi)-2\alpha^2 \Im(\hat{\psi})\mathfrak{I}(\ddey\hat\psi),
\end{align}
where $\Re, \Im$ stand for real and imaginary part, respectively, and the integral enstrophy as
\begin{align}\label{eq:enstrophy}
\Omega&=\nonumber\frac{1}{4}\int^1_{-1} \|\hat\omega\|^2 \ud y \\&=\frac{1}{4}\int^1_{-1} \left(\|\ddey\hat\psi\|^2+2\alpha^2\|\dey\hat\psi\|^2+\alpha^4\|\hat{\psi}\|^2\right)\ud y.
\end{align}

%

It is now interesting to observe that the integral enstrophy can be split in two parts:
\begin{align}\label{eq:enstrophy1}
\Omega&=\alpha^2E+F,
\end{align}
where
\begin{align}
\label{eq:eqE}
E&=\frac{1}{4}\int^1_{-1} (\|\dey\hat\psi\|^2+\alpha^2\|\hat\psi\|^2)\ud y,
\end{align}
is the integral kinetic energy of the perturbation, and  
\begin{align}
F&=\frac{1}{4}\int^1_{-1}( \|\ddey\hat\psi\|^2+\alpha^2\|\dey\hat{\psi}\|^2)\ud y,
\end{align}
a positive quantity related to the streamwise component of the velocity perturbation and its cross-derivative. Notice that, for wavenumbers equal or greater than one, the integral enstrophy is always greater than the integral kinetic energy. Wavenumbers of order one are typically the most unstable, both asymptotically and in the transient \cite{cjj2003,schmid2007}. Notice also that in the limit $\alpha\to 0$ the integral enstrophy is independent on the transversal perturbation velocity.

The temporal evolution equations for $E$ and $F$ are derived as follows:
\begin{align}
\dtot{E}{t}&=-\mathfrak{R}\Bigl\{ \frac{1}{4}\int^1_{-1}\Bigl[\bar{\psi}\ \dt(\ddey\hat\psi-\alpha^2\hat{\psi})\Bigr]\ud y\Bigr\}\nonumber\\
&=-\mathfrak{R}\Bigl\{\frac{1}{4}\int^1_{-1}\Bigl[\bar{\psi}\ \Bigl(   -i\alpha U \ddey \hat{\psi}+i\alpha^3U\hat{\psi}+i\alpha U''\hat{\psi}\nonumber\\&
+Re^{-1}(\ddddey \hat{\psi}-2\alpha^2 \ddey \hat{\psi}+ \alpha^4 \alpha^2 Re^{-1})    \Bigr)\Bigr]\ud y\Bigr\}, \label{eq:dEdt}\\
\dtot{F}{t}&=-\mathfrak{R} \Bigl\{\frac{1}{4}\int^1_{-1}\left[\ddey\bar{\psi}\ \dt(\ddey\hat\psi-\alpha^2\hat{\psi})\right]\ud y\Bigr\}\nonumber	\\ 
&=\mathfrak{R}\Bigl\{\frac{1}{4}\int^1_{-1}\Bigl[\ddey\bar{\psi}\ \Bigl(   -i\alpha U \ddey \hat{\psi}+i\alpha^3U\hat{\psi}+i\alpha U''\hat{\psi}\nonumber\\&
+Re^{-1}(\ddddey \hat{\psi}-2\alpha^2 \ddey \hat{\psi}+ \alpha^4 \alpha^2 Re^{-1})    \Bigr)\Bigr]\ud y\Bigr\},\label{eq:dFdt}
\end{align}
where the upper bar stands for complex conjugate and $\mathfrak{R}$ for real part. At this point, the enstrophy equation for a small wavy perturbation can be obtained as $\ud(\alpha^2E+F)/\ud t$ from the two equations above. By considering the basic flow expressions and the boundary conditions, we obtain
\begin{align}\label{eq:dtenstrophy_testo}
\nonumber
\dtot{\Omega}{t}=& Re^{-1}\mathfrak{R}\Bigl[\dddey\hat{\psi} \ddey\bar{\psi}\Bigr]^1_{-1}\\\nonumber&-{Re}^{-1}\int^1_{-1} \big(3\alpha^2 \|\ddey\hat{\psi}\|^2+3\alpha^4\|\dey\hat{\psi}\|^2\\&+\alpha^6\|\hat{\psi}\|^2+\|\dddey\hat{\psi}\|^2\big)\ud y={Re}^{-1} H.
\end{align}
It should be noted that the three convective terms in Eq. (\ref{eq:OrrSomm}), which contain as factors the basic flow $U$ and its second derivative $U''$, do not appear in the above equation. This is due to the canceling of some terms in  Eq. (\ref{eq:dEdt}) and ( \ref{eq:dFdt}), which takes place when the real part is taken.
Other terms vanish since they are contained in both $\alpha^2E$ and $F$ with opposite sign. As a consequence, the temporal enstrophy evolution  is physically determined by the diffusive terms of the motion equation only, and  $Re^{-1}$ can be factored out. On top of that, \emph{it is of great interest that the only term which can generate a temporal growth of enstrophy is the boundary term associated with the wall vorticity and its cross-flow variation at the walls}.
\par

The aim of the present study is to find the exact lower bound $Re_{\Omega}$ for the enstrophy transient growth of any 2D perturbations. That is, we look for
\begin{equation}\label{eq:ReZ} Re_\Omega(\alpha; U(y))=\sup_{\hat\psi(y,t=0)}\left\{ Re \hskip5pt  : \hskip5pt \dtot{}{t}\Omega\leq 0,\hskip5pt \forall t\right\}, \end{equation}  
meaning that for $Re>Re_\Omega$ there exists at least one initial condition leading to a temporal enstrophy growth in the transient. When  $Re<Re_\Omega$ instead, the enstrophy of any initial perturbation can only experience  monotonic time decay. \par
It is interesting to focus on the term $\mathfrak{R}[\dddey\hat{\psi} \ddey\bar{\psi}]^1_{-1}$ in Eq. (\ref{eq:dtenstrophy_testo}), since it is the only term which can be positive and can thus induce a possible growth. 
%
However, boundary conditions on the vorticity are notoriously unknown \textit{a priori}, as first underlined by Synge in 1935 \cite{synge1935}. This fact has represented the main obstacle to the solution of problem (\ref{eq:ReZ}).

The mathematical formulation developed by Synge in 1938 was a peculiar application of the modal temporal theory to the vorticity equation, see Eq. 11.28 in \cite{synge1938a} and Eq. 2.5 in \cite{synge1938b}. In synthesis, the method is the following. By multiplying the OS equation by  $e^{\pm\alpha y}$ and integrating, Synge obtained the following two integral relationships,
\begin{equation}\label{eq:vorticity_dynamical_testo}
	\left[(\dddey\hat{\psi}\pm\alpha \ddey \hat{\psi})e^{\alpha y}\right]_{-1}^1=\mp 2i\alpha^2 Re\int^1_{-1} U'\,\hat{\psi} e^{\alpha y}\ud y,
\end{equation}
 which link the  wall values of the vorticity and its y-derivative (actually, the part of the vorticity  associated with the cross-flow momentum variation). Such ``dynamical condition'' has to be satisfied by the streamfunction, as discussed by Synge in 1935 \cite{synge1935}. By using the above expressions, he wrote un updated integral enstrophy equation which was then optimized to maximize the enstrophy time variation as a function of $Re$. 
At the time, the author aimed at finding a lower bound for linear asymptotic stability and, ultimately,  conditions for linear instability. That is, the focus was on seeking the unconditional instability threshold $Re_\mathrm{c}$ (shown in Fig. \ref{fig:maps_summary}(\textbf{b}), for the plane Poiseuille flow). This justified the use of the exponential time factor in the perturbative hypothesis. Today, we know that $Re_\mathrm{c}=5772$ \cite{orszag1971} for PPF, while   $Re_\mathrm{c}=\infty$ \cite{romanov1973} for PCF. Since at the time the phenomenon of non-modal transient growth was unknown  \cite{ttrd1993}, Synge could not be aware that his computations would lead  to a much lower bound for the algebraic transient growth of the vorticity. His calculations worked out for the plane Poiseuille flow but not for the plane Couette flow, where symbolic calculus helped us to accomplish the task. In place  of the exponential time dependence $\psi=\hat{\psi}(y) e^{i \alpha x - \sigma t }$, we use the non-modal approach where $\psi=\hat{\psi}(y,t) e^{i\alpha x}$ and solve Eq. (\ref{eq:ReZ}) for both flows.

\section{Results and Discussion}\label{sec:limiting_curves}

The  complete mathematical procedure developed in order to solve the problem (\ref{eq:ReZ}) is given in the Appendix. In the following, the main steps are recalled. Even though our procedure does not impose the analytical temporal structure of solutions, conditions for enstrophy monotonic decay in the Poiseuille case have been formally derived as done by J. L. Synge in \cite{synge1938b}. 

The route to the solution of the problem (\ref{eq:ReZ}) is made of four main steps: \\(i) Derive the conditions (\ref{eq:vorticity_dynamical_testo}).\\
\textcolor{black}{  (ii) Use Eq. (\ref{eq:vorticity_dynamical_testo}) in the enstrophy equation (\ref{eq:dtenstrophy_testo}). Obtain the enstrophy growth rate, $\ud \Omega/\ud t$, parametrized with  the possible boundary terms $\ddey \hat\psi(\pm1,t)$:
 \begin{align}
 \nonumber
 \dtot{}{t}\Omega=&{Re}^{-1} H={Re}^{-1} \Bigl[ \alpha a b^{-1}\Theta-\alpha b^{-1} \Phi+i\alpha^2 Re\,b^{-1}B+\\&-(I_3^2+3\alpha^2I_2^2+3\alpha^4I_1^2+\alpha^6I_0^2)\Bigr],
 \end{align}
 where, using the bar symbol for the complex conjugate,
 \begin{align}
 I_i^2=&\int^1_{-1} \partial_y^{(i)}\hat{\psi}  \partial_y^{(i)}\bar\psi \ud y,\\
 \Theta= \,&\ddey\hat{\psi}(1,t)\ddey\bar{\psi}(1,t)+\ddey \hat{\psi}(-1,t)\ddey\bar{\psi}(-1,t) ,\\
 \Phi= \,&\ddey\hat{\psi}(1,t)\ddey\bar{\psi}(-1,t)+ \ddey\hat{\psi}(-1,t)\ddey\bar{\psi}(1,t),\\
 B=& \int_{-1}^1 \Bigl[\hat{\psi}(y,t)\ddey\bar{\psi}(1,t)-\bar{\psi}(y,t)\ddey\hat{\psi}(1,t)\Bigr]U'\nonumber\\&\times\cosh[\alpha(y+1)]\ud y-\int_{-1}^1\Bigl[\hat{\psi}(y,t)\ddey\bar{\psi}(-1,t)\nonumber\\&
 -\bar{{\psi}}(y,t)\ddey\hat{\psi}(-1,t)\Bigr]U'\cosh[\alpha(y-1)]\ud y.
 \end{align}} 
 
\noindent \textcolor{black}{(iii) By calculus of variations, obtain the following 6th-order PDE for the perturbation $\hat\psi_m(y; t)$ which maximizes the enstrophy growth rate:
\begin{align}
\nonumber
\partial_y^6  \hat{\psi}_m(y,t)-&3\alpha^2 \ddddey \hat{\psi}_m(y,t)+3\alpha^4 \ddey  \hat{\psi}_m(y,t)-\alpha^6 \hat{\psi}_m(y,t)\\
\nonumber
=&i\alpha^2 Re\,b^{-1}U'(y)\Bigl\{\ddey\hat{\psi}_m(1,t)\cosh[\alpha(1+y)]\\\label{eq:sixthorderpsi_testo}
&-\ddey\hat{\psi}_m(-1,t)\cosh[\alpha(1-y)]\Bigr\}.
\end{align}}
  	\begin{figure}[]
  		\centering
  		\includegraphics[width=.96\columnwidth]{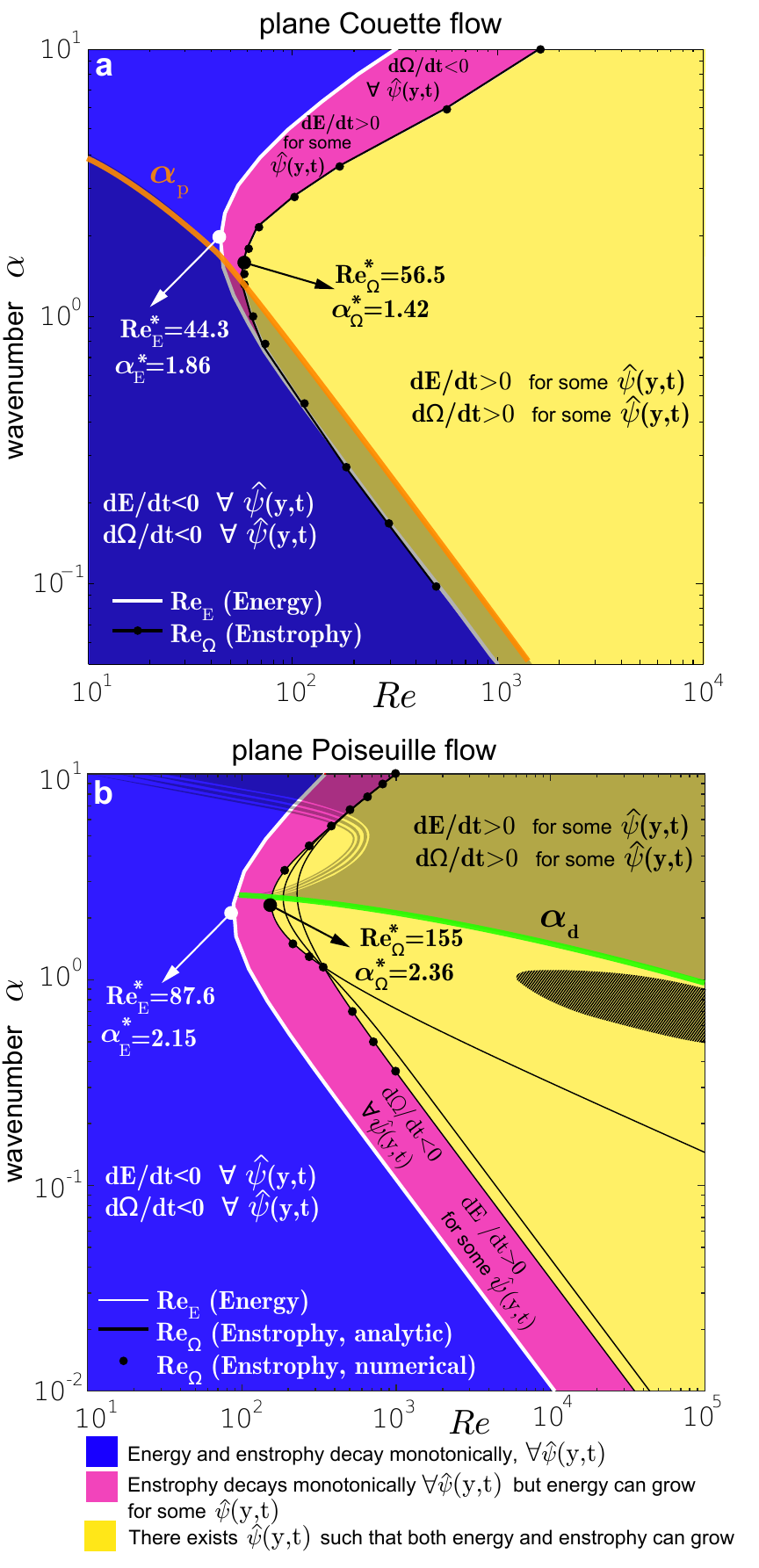}
  		\caption{\textbf{Lower bounds for the transient growth of enstrophy and  kinetic energy of  2D waves. Propagation properties.} \textit{Blue region}:  both the kinetic energy  ($E$) and the enstrophy ($\Omega$)  decay monotonically with time, regardless of the initial condition. \textit{Pink region}: transient kinetic energy growth is possible, but enstrophy growth is not. \textit{Yellow region}: growth of both kinetic energy and enstrophy is allowed.  \textbf{(a)} PCF. In this case, the curve $Re_\Omega(\alpha)$ (black dotted) was computed numerically via an optimization procedure.   The shaded region indicates where wave propagation is forbidden \cite{gallagher1962}. \textbf{(b)} PPF. Here $Re_\Omega(\alpha)$ was computed both analytically (black curves), \textcolor{black}{see Eqs. (\ref{eq:QuattroUndici}), (\ref{eq:QuattroTredici}}), and numerically (black dots). In the shaded region waves are non-dispersive in the long term, while different levels of dispersion are observed in the remaining part of the map \cite{desanti2016} (notice both the sharp lower boundary $\alpha_\mathrm{d}$  and the smooth transition in the upper part). In black, we show the unconditional instability region \cite{orszag1971}, not existing in the PCF case \cite{romanov1973}.}\vskip-10pt
  		\label{fig:maps_summary}
  	\end{figure}
\noindent(iv) Solve Eq. (\ref{eq:sixthorderpsi_testo}) and obtain, from the corresponding maximal enstrophy functional, the region of the $\alpha-Re$ map where transient enstrophy growth is not allowed, that is the curve $Re_\Omega(\alpha)$. \\
This final curve was computed both analytically and via numerical optimization for PPF, just numerically for PCF (see Appendix). \textcolor{black}{Both the analytical inequalities and the numerical optimization are aimed at finding the best solution over the possible values of the vorticity at the wall, $\ddey\hat{\psi}(\pm1,t)$}.

The results of our calculations are shown in Fig. \ref{fig:maps_summary}(\textbf{a},\textbf{b}).
The minimum value of $Re_\Omega$ for which 2D perturbations can experience transient enstrophy growth is named $Re^*_{\Omega}$. In the case of plane Couette flow, we found the value $56.5$, at a wavenumber $\alpha^*_{\Omega}=1.42$. For the plane Poiseuille flow  $Re^*_{\Omega}=155$ at $\alpha^*_{\Omega}=2.36$ (see Fig. \ref{fig:maps_summary}).  

\textcolor{black}{The shape and of enstrophy-rate optimal stream-functions for ($Re$, $\alpha$) close to the marginal curve $Re_\Omega$ are shown in Fig. \ref{fig:marginal_optimal} in the Appendix. The procedure used to compute these solutions is also described in details in the Appendix. In the Supplemental Material we also report maximizing solutions in both the monotonic decay region and in the transient growth region of the stability map, together with their temporal evolution (see Fig. SM 2)}.
	
Figure \ref{fig:maps_summary} also compares the enstrophy lower bound for transient growth with the bound for the kinetic energy transient growth, that is the curve $Re_\mathrm{E}(\alpha)$. The kinetic energy problem was first formulated and solved by Orr \cite{o1907a}, and subsequently by Synge \cite{synge1938a} and Joseph \cite{joseph1976}, while numerical solutions for the three-dimensional case have been obtained later on by Reddy \& Henningson \cite{rh1993}.


We computed $Re_\mathrm{E}$ using the energy method, based on a variational formulation \cite{o1907b,joseph1976,rh1993}. This bound is represented by the white curves in Fig. \ref{fig:maps_summary} and, in Fig. \ref{fig:mappa_massimi}(\textbf{a},\textbf{b}), by the right boundary of the white regions.

A relevant outcome of our analysis is that the threshold for enstrophy monotonic decay $Re_\Omega(\alpha)$ for 2D waves is greater than the threshold for the kinetic energy $Re_\mathrm{E}(\alpha)$, at any wavenumber. The gap between the two is highlighted by the pink region of Fig. \ref{fig:maps_summary}. This means that there exists a region in the $\alpha$-$Re$ space where transient kinetic energy growth can occur, while enstrophy growth is forbidden, for any initial perturbation. \par

\begin{figure}[t]
	\centering
	\includegraphics[width=\columnwidth]{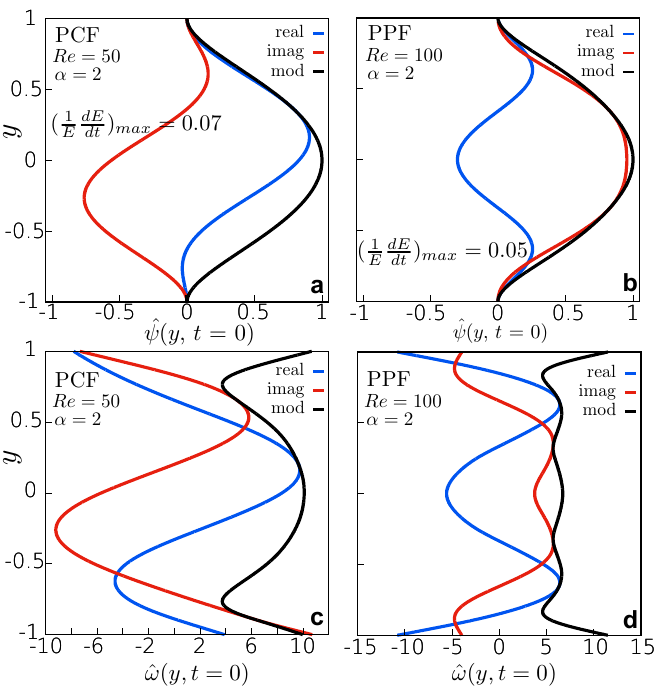}\vskip-10pt
	\caption{\textbf{Energy-rate optimal initial conditions used to build the map of  Fig. \ref{fig:mappa_massimi}}.    \textbf{(a)} PCF. This perturbation  maximizes the initial kinetic energy growth rate at $Re=50$ and $\alpha=2$. \textbf{(b)} PPF. In this case, the initial condition maximizes the energy growth rate at $Re=100$ and $\alpha=2$. Such perturbations excite the least stable Orr-Sommerfeld eigenfunctions, and contain both symmetric and antisymmetric modes. Panels \textbf{(c)} and \textbf{(d)} show the  shape of the corresponding initial vorticity, $\widehat\omega=\alpha^2\widehat\psi-\partial^2_y\widehat\psi$.}\vskip-10pt
	\label{fig:init_conditions}
\end{figure}

\begin{figure}
	\centering
	\includegraphics[width=\columnwidth]
	{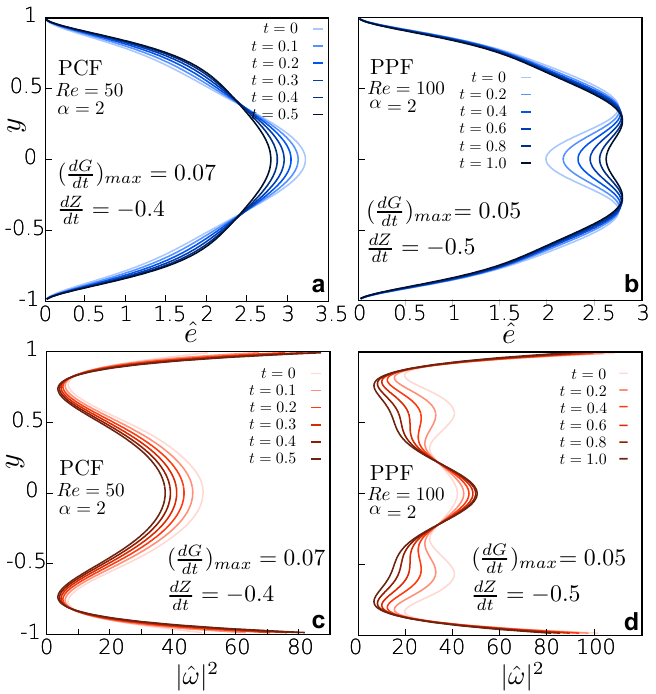}
	\caption{\textcolor{black}{\textbf{Instances of simultaneous kinetic energy growth and enstrophy decay}. For the two initial conditions of Fig. \ref{fig:init_conditions}, this figure shows the temporal behavior of kinetic energy y-profiles (top panels) $\hat e(y,t)=\frac{1}{2}(\| \dey\hat\psi \|^2+\alpha^2\| \hat\psi\|) $ and local enstrophy distribution (bottom panels), during the time interval along which the integral kinetic energy experiences a transient growth. Left panels show PCF, right panels PPF. 
    The maximal kinetic energy rate  and the enstrophy rate (at the initial instant, $t=0$) are reported in all panels ($G=E/E_0$,  $E(t)=\int_{-1}^1 \hat e(y,t)\ud y$; $Z=\Omega/\Omega_0$). }
	}\label{fig:norm_ene_ens}
\end{figure}

\begin{figure*}[t]
		\centering
		\includegraphics[width=0.77\textwidth]{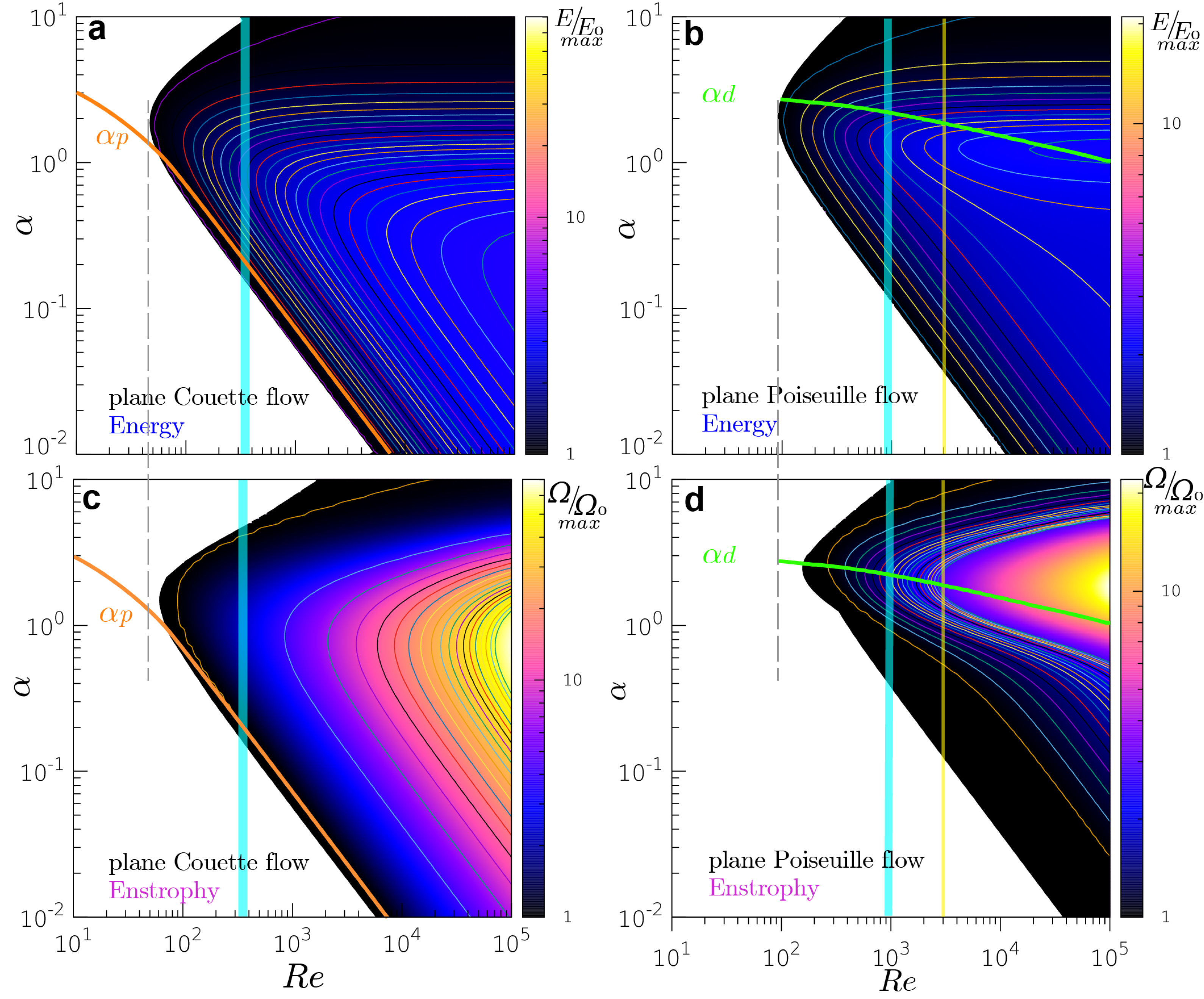}\vskip-10pt
		\caption{\textbf{Maximal transient growth of perturbation enstrophy and kinetic energy. A case study.}\\  Wavenumber -- Reynolds number maps of maximal transient growth of kinetic energy ($E/E_0$, top panels \textbf{a}, \textbf{b}) and enstrophy ($\Omega/\Omega_0$, bottom panels \textbf{c}, \textbf{d}), normalized to the initial value. Left panels regard the plane Couette flow (PCF), right panels the plane Poiseuille flow (PPF). Each map is built from 3600 numerical simulations of the initial-value problem associated with Eq. (\ref{eq:OrrSomm}) (60 values of $\alpha$ in the range $[10^{-2},10]$ and 60 values of Re  in $[10,10^5]$, uniformly distributed in the log space). The initial condition is shown in Fig. 	\ref{fig:init_conditions}(\textbf{a}) and \ref{fig:init_conditions}(\textbf{b}) for PCF and PPF, respectively. Contours start from  $1.01$ and their spacing is set to 0.1 in panels  \textbf{(a)}, \textbf{(b)}, \textbf{(d)}, while levels spacing is set to 3 in panel \textbf{(c)}.	
		The light-blue vertical bands represent the \textit{global stability} 3D threshold $Re_\mathrm{g}$: values collected from experiments in the literature are around 325 for PCF and 840 for PPF, and are here reported for comparison reasons. The bands width stands for the range of values found in the extensive literature on the subcritical transition to turbulence \cite{orszag1980,carlson1982,lundbladh1991,daviaud1992,ttrd1993,waleffe1995a,waleffe1995b,bottin1998,bottin1998b,duguet2010,manneville2012,tsukahara2014,manneville2015a}. In 2D, nonlinear analysis  of PPF leads to a transitional value of about 2900 \cite{bayly1988} (vertical yellow line), while for PCF no results are yet available. All maps include information on wave propagation and dispersion. The green curve  in  PPF panels represents the threshold $\alpha_\mathrm{d}(Re)$, between dispersive and non-dispersive longterm behavior (below and above the curve, respectively, see  \cite{desanti2016}). In the PCF case instead, below the orange curve $\alpha_\mathrm{p}(Re)$  waves are stationary \cite{gallagher1962}.}\vskip-10pt
		\label{fig:mappa_massimi}
	\end{figure*}

\textcolor{black}{This finding can be seen as counter-intuitive as we typically envision perturbations in their temporal asymptotic state (exponentially growing or decaying waves). In the far term indeed, the normalized shape of the perturbation is not varying any more ($\hat\psi(y,t)/\|\hat\psi(y,t)\|_\infty=f(y)$), and consequently the kinetic energy and the enstrophy must follow the same trend.  The dynamics is different during the early-transient evolution: here, the perturbation is changing shape and cancellation effects among non-normal OS modes allow for transient growth. However, it is possible that the integral kinetic energy grows while the volume enstrophy does not.  Two examples of this kind of perturbations are shown in Fig. \ref{fig:init_conditions}. We used such perturbations  as initial conditions for the numerical simulations which are used to build the maps of Fig. \ref{fig:mappa_massimi} and Fig. SM 1. These perturbations guarantee a positive kinetic energy growth rate at $t=0$ (at $Re=50,\alpha=2$ for PCF and $Re=100, \alpha=2$ for PPF) but no enstrophy growth can be observed. Figure \ref{fig:norm_ene_ens} shows the temporal evolution of such perturbations in terms of y-distribution of kinetic energy (top panels) and enstrophy (bottom), during the time interval when the kinetic energy experiences a transient growth and the enstrophy decays. In these two particular cases, the kinetic energy increases near the channel center for PPF and close to the walls for PCF. The same qualitative behavior is found for the enstrophy, but here the growth is smaller and located at comparatively narrower regions across the channel, a fact that produces the decay of the integral enstrophy.}

\textcolor{black}{In addition to the results given by the analytical procedure, by using the numerical method described in reference \cite{desanti2016}, we performed numerical simulations of the initial-value problem (\ref{eq:OrrSomm}). Wavenumber-Reynolds maps of the maximal enstrophy and kinetic energy reached in the transient evolution are shown in Fig. \ref{fig:mappa_massimi}. To our knowledge, enstrophy maps have not yet been presented in the literature. Instead, maps of kinetic energy have been previously shown \cite{g1991,rsh1993,schmid2007}. They typically present the maximum amplification over all  possible initial conditions. Here, we follow a different approach where we keep the initial condition fixed. The initial condition satisfies general features of smoothness, excitation, of both symmetric and antisymmetric Orr-Sommerfeld modes (Fig. \ref{fig:init_conditions}), and it was chosen in order to trigger  a transient energy growth for any $Re$ above the limit $Re^*_\mathrm{E}$. From an optimization process, we got the perturbation leading to the maximal kinetic energy growth rate in the surrounding of the map ``nose'' ($\alpha^*_\mathrm{E}$, $Re^*_\mathrm{E}$). As predicted by the analytical result, the vorticity starts to experience a transient growth  for $Re(\alpha)>Re_\Omega(\alpha)$ only, see Fig. \ref{fig:mappa_massimi}(\textbf{c}, \textbf{d}).}

\textcolor{black}{Comments are now proposed about the map structure.  It is observed that the internal structure of both enstrophy and kinetic energy maps reflects the shape of the respective lower bound for transient growth, see level curves in Fig \ref{fig:mappa_massimi}. This feature is clearly seen in the low-wavenumber region, and can be interpreted as the scaling laws  $\Omega_{max}\sim(\alpha Re)^{\delta_1}$, $E_{max}\sim(\alpha Re)^{\delta_2}$. The exponents $\delta_1$, $\delta_2$ depend on the initial condition, and for the cases observed here: $\delta_1\approx0.82$ for PCF, $\delta_1\approx0.21$ for PPF; $\delta_2\approx0.59$ for PCF, $\delta_2\approx0.33$ for PPF (computed for $\alpha<0.1$). Inside the region of the map were both the wave kinetic energy and the enstrophy can grow, we observe that smooth vortical initial disturbances show a comparatively much higher amplification of integral enstrophy than kinetic energy. Furthermore, when the non-dimensional time necessary to achieve the maximal growth is considered,  a unique scaling is observed for the enstrophy and the kinetic energy in the high-$Re$ and low-$\alpha$ limit, as shown in Fig. SM 1: $T_{E_{max}}\sim T_{\Omega_{max}}\sim\alpha^{-1}$.} \par

\textcolor{black}{Algebraically amplified waves are located inside a nearly conical region of the map with the apex towards wavenumbers  $\alpha\sim 2$, see Fig \ref{fig:mappa_massimi}.  This trend generally holds in the three-dimensional case \cite{rh1993,ttrd1993,schmid2007}. Notice also that for PPF the exponential growth is found at $\alpha \approx [ 0.3 - 1]$). In this study, however, we thought important to extend the range of observed wavenumber and Reynolds number. Let us remind that in  subcritical conditions ($Re<Re_c$) the transition is triggered by spatially localized perturbations.  A local perturbation can be described as a wave packet which typically contains a broad range of wavenumbers. We believe that at least two physical factors concur to the onset of nonlinear interaction in such situations. The first is the algebraic growth of kinetic energy or, better, enstrophy.  The second is related to the dispersion of the wave components.} \textcolor{black}{In fact, beside the wave amplification, the propagation properties play a key role in the nonlinear coupling onset. As we showed in a previous study on the plane Poiseuille and wake flows (3D, see Fig. 2 in \cite{desanti2016} and Chapter 2 in \cite{fraternale_phdthesis}), wave dispersion can be significantly different for large and small wavenumbers, and \textit{wave dispersion and non-dispersion coexist within the same flow}. For PPF, there exists a dispersive-to-nondispersive transition of the least-stable mode, which occurs at a specific wavenumber ($\alpha_\mathrm{d}$, in the following). 
Here, the results of our previous study are extended to a Reynolds number range of four-decades.} We measure the dispersion intensity in terms of the difference between the non-dimensional group velocity and phase velocity.  It can be noticed that the parameters space is split in regions having different dispersion characteristics.  \textcolor{black}{The discriminant wavenumber $\alpha_\mathrm{d}$ is represented by the green curve in Fig. \ref{fig:maps_summary}(\textbf{b}), and in  Fig. \ref{fig:mappa_massimi}(\textbf{b},\textbf{d}).}
Below this boundary, the waves travel dispersively,  \textcolor{black}{slower than the basic flow $U_\textrm{CL}$, and have large vorticity close to the walls.
Oppositely, the motion of short waves with $\alpha>\alpha_\mathrm{d}$ is convective and the behavior is mostly nondispersive, in particular for $Re>1000$, (see the shaded region of Fig. \ref{fig:maps_summary}\textbf{b}). In this case the largest vorticity is located at the channel center. Even if short-waves growth is mild, they are responsible for the generation of compact structures that have been observed in laboratory and numerical experiments}. 
\begin{figure}[h]
	\centering
	\includegraphics[width=0.85\columnwidth]{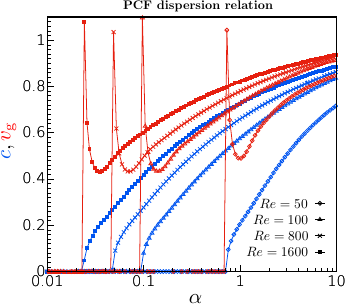}
	\caption{\color{black}\textbf{Dispersion of the least-damped mode for the plane Couette flow}. Distribution of the phase velocity $c$ (light blue) and the group velocity $v_\mathrm{g}=\ud\omega/\ud\alpha$ (red) are shown for $\alpha$ in the range $[0.01-10]$ for the least-damped Orr-Sommerfeld mode in the plane Couette flow for $Re=[50,100,800,1600]$. The computation was performed by a 4th-order finite difference scheme \cite{desanti2016}, the wavenumber  is uniformly discretized in the log-space (1024 points). For clarity, for each curve, only one every ten points is shown. We remind that since the three-branched Orr-Sommerfeld eigenvalues spectrum of PCF is symmetric about the frequency axis, there always are two modes equally damped, traveling in opposite directions.  Below the threshold $\alpha_\mathrm{p}(Re)$, wave propagation is forbidden. At higher wavenumbers, wave dispersion is observed: it is high in the neighborhood of  $\alpha_\mathrm{p}$ and it decreases as $\alpha Re\to \infty$.}
	\label{fig:dispersion_couette}
\end{figure} 
For the plane Couette flow, such an abrupt transition between dispersive and non-dispersive behavior does not exist, since small traveling waves always disperse. The dispersion is mild, but it becomes intense close to a boundary curve that we call  $\alpha_\mathrm{p}(Re)$, below which waves become stationary  (Fig. \ref{fig:dispersion_couette} shows the dispersion relation of PCF). The bound $\alpha_\mathrm{p}$ is represented by the orange curve in Fig. \ref{fig:maps_summary}(\textbf{a}), Fig. \ref{fig:mappa_massimi}(\textbf{a},\textbf{c}) and Fig. SM 1. Such threshold was first found by Gallagher \& Mercer in 1962 \cite{gallagher1962}. 
 \begin{figure*}[]
 	\centering
 	\includegraphics[width=0.48\textwidth]{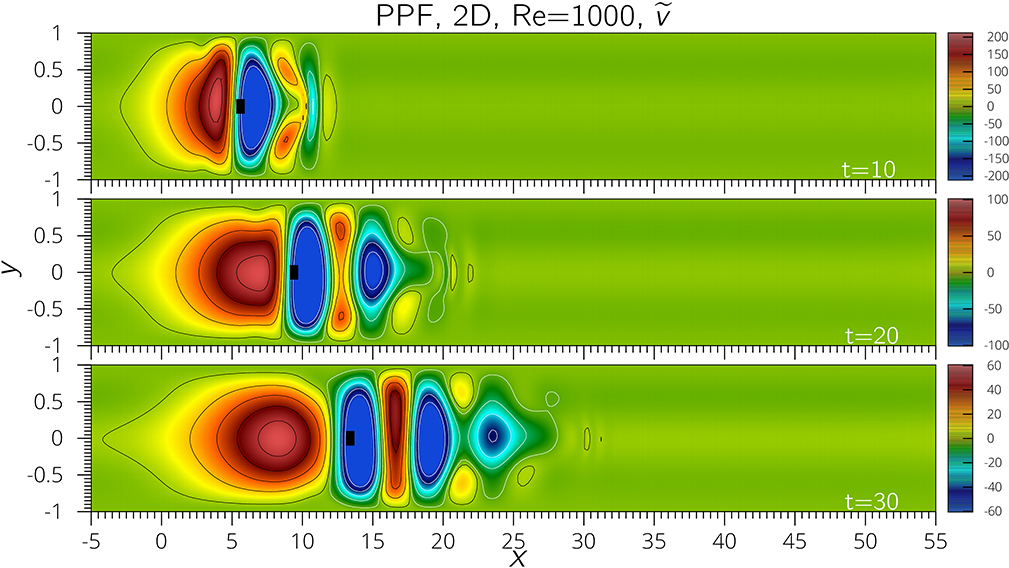}
 	\includegraphics[width=0.48\textwidth]{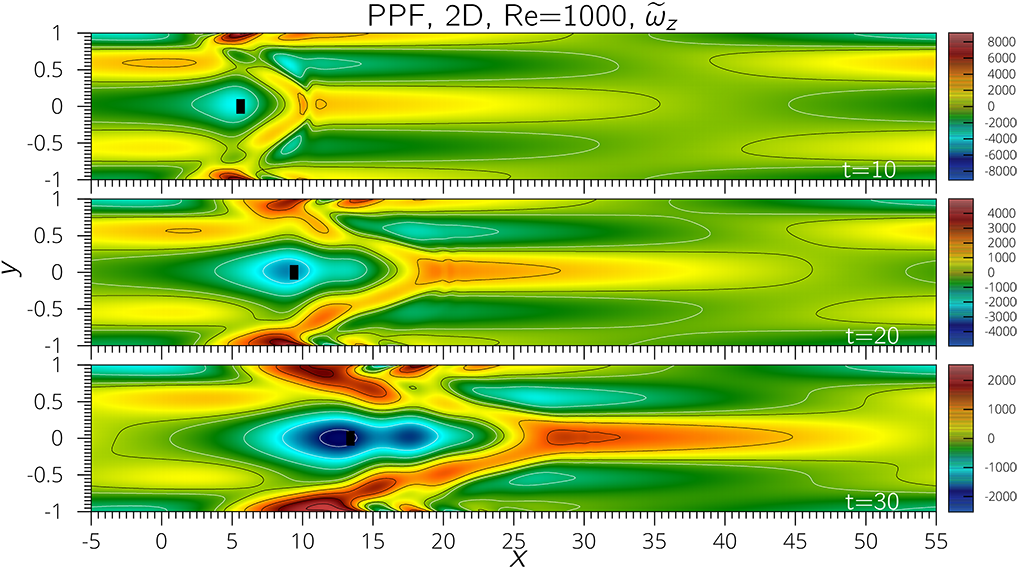}
 	\caption{\color{black}\textbf{Temporal evolution of a linear 2D wave packet in plane Poiseuille flow at Re=1000}. The packet consists of 512 waves with wavenumber in the range $[0.1-10]$ In the left panels, the  $\widetilde v$ disturbance evolution is shown  at $T=10, 20, 30$ (top to bottom). In the right panels, the corresponding perturbation vorticity is visualized. The basic flow is from left to right. The initial perturbation is localized at $x_0=0$. Its y-distribution is that of Fig. \ref{fig:init_conditions}, while the x-distribution is chosen to be Gaussian (standard deviation is 5\% of the channel width). The initial peak value is $\widetilde v_\mathrm{max}(t=0)=1530$, (irrelevant to the packet evolution, in the linear context). For each panel, five contour lines are traced, the first level is $0.05\ \widetilde v_{max}$ and the last one is $0.9\ \widetilde v_{max}$, where the peak value  $\widetilde v_{max}$ can be inferred from  the color bars. The dynamics shows a fast, compact, front moving with the centerline speed of the basic flow, and a  slower rear part. We address the former to the non-dispersive range of wavenumbers ($\alpha\sim[2,10]$), while the spot core is related to the dispersive wave components in the lower part of the stability map. Both components contribute to the formation of a shear layer which is characteristic of subcritical flow structures \cite{breuer1988,breuer1990,landahl1975,klingmann1992,henningson1993,kerswell2005}. This figure is taken from F. Fraternale's PhD dissertation \cite{fraternale_phdthesis}.  }
 	\label{fig:2Dspot_Re1000}
 \end{figure*}
 
From here it is possible to infer that any spatially localized perturbations (wave packets), which contain a broad range of traveling wave components,  can present both the dispersive and the non-dispersive behavior. There will be a subset of dispersive waves that spread information in the surrounding environment, enhancing the probability to intercept other neighbor perturbations. In case the enstrophy is sufficiently amplified (see figure \ref{fig:mappa_massimi}), this could trigger a nonlinear coupling.
On the other hand, packets include also a non-dispersive subset of waves which are propagate as the basic flow.  Once again, if the enstrophy and kinetic energy content is sufficiently high the onset of a nonlinear coupling can be expected, since this subset does not unpack.  
 
\textcolor{black}{Notice that similar scenarios have also been observed in pipe flows \cite{avila2011}. In the context of liquid films, as well, the flow stability was found to be significantly affected by wave dispersion, by phase-synchronization of stable modes yielding to generation of ``explosive disturbances'' \cite{benilov2003,benilov2004,craik2005}.} 

\textcolor{black}{In a natural context of transition onset, it unlikely to observe individual waves. Usually instead, wave packets are observed.  As said in the previous paragraphs, the morphology of such packets depends on both the	growth rates of enstrophy and energy and on the dispersion properties associated with the individual waves contained therein. We  show the results of a numerical simulation of a 2D, localized, linear wave packet in the plane Poiseuille flow (wall-normal velocity and vorticity are visualized in Fig. \ref{fig:2Dspot_Re1000}(\textbf{a}, \textbf{b})).
From the vorticity visualization, it is possible to notice the intense shear layer which is typically observed in dynamics of subcritical transitional flow structures.  This layer was first observed in the 1980s in boundary layers by Breuer \& Haritonidis \cite{breuer1990}, Breuer \& Landahl \cite{breuer1990b}, and in PPF by Klingmann \cite{klingmann1992} and Henningson, Lundbladh \& Johansson \cite{henningson1993} (see, for instance Fig. 6 and Fig. 18 in \cite{klingmann1992}, and Fig. 3 in \cite{henningson1993}). Similar structures are also peculiar of puffs in pipe flows, see for instance \cite{wygnanski1973,duguet2010b,avila2011}. Such shear layer has a typical lambda-shape heading downstream, made of two layers which take origin at the walls in the slow dispersive region of the packet and merge at the channel centerline, generating the spot's front. The physical mechanism leading to this structure is the \textit{lift-up effect} described by Landahl in 1975 \cite{landahl1975}. 
Although this mechanism is mostly three-dimensional (is mainly related to the ``vortex tilting term'' $i\beta U' \widehat v$ which appears at the right hand side of the Squire equation), it is also present in the 2D case. In the two-dimensional case indeed, the mechanism is related to the $i\alpha U''\hat\psi$ term of the Orr-Sommerfeld equation.  Landahl discovered the formation of an elongated \textit{permanent scar} convected downstream with the local basic flow speed. The term \textit{permanent} refers to the much longer decay time experienced by this streamwise perturbation, with respect to that of the wall-normal perturbation $\widetilde v$. For this reason, the shear layer is not visible from the $\widetilde v$ component but is can be observed from the streamwise velocity $\widetilde u$ or from the spanwise vorticity. 
 }

\section{Conclusions}	 
For 2D, viscous, vortical internal waves in the plane Couette and Poiseuille flows, we determined the exact lower bound for the enstrophy transient growth.  This result was obtained following J.L. Synge's approach (1938), which at that time had already been conceived as an alternative to kinetic-energy based analysis.
			
 As far as the monotonic decay is concerned, it is found that at all  wavenumbers this bound is less restrictive than the kinetic energy one,  that is $Re_{\Omega}  (\alpha) > Re_\mathrm{E} (\alpha)$. This is physically noticeable, it is indeed not intuitive that an initial vortical perturbation which experiences a quick kinetic energy growth does not necessarily experience  an enstrophy transient growth.

\textcolor{black}{Our study provides new maps for the maximal perturbation enstrophy growth and related time scales. In the low-wavenumber part of the parameters space, this yields information on the scaling law for both the enstrophy and kinetic energy maximal growth. We highlight that Poiseuille and Couette maps differ more in the enstrophy case rather than in the kinetic energy case. }

In addition, by building on the results of \cite{desanti2016}, we  underline  the  notable variability of the dispersion properties within the parameters space. At a fixed $Re$, by moving inside the parameters space from very low wavenumbers, in the Couette case one can pass from stationary waves to dispersive waves which then become progressively less dispersive, reaching a quasi-convective propagation. In the Poiseuille case instead, one can pass abruptly from quite dispersive waves to non-dispersive waves across the curve $\alpha_\mathrm{d}$.
\textcolor{black}{Within a spatially localized perturbation as a wave packet, both a dispersive and a non-dispersive subset of waves can be present. 
Generally, a wave packet will be composed by both a dispersive subset of waves which spread out the disturbance on a larger portion of the spatial domain and by a non-dispersive subset of waves which travel in a compact fashion.  The inference can be made that in turn, or also  simultaneously, both these components may contribute to the nonlinear wave coupling when a sufficient enstrophy amplification, rather than a kinetic energy one, is reached.}


\acknowledgments{The authors acknowledge support from the US NSF (grants DMS 1362509 and DMS 1462401) and from the MISTI-Seeds Italy MITOR project ``Long-term interaction in fluid systems", 2012-2014, \url{http://web.mit.edu/mitor/grants/seed.html}. Computational resources were provided by HPC@POLITO (\url{http://www.hpc.polito.it}).}

\appendix

\FloatBarrier

\section{Mathematical procedure to obtain the maximum time derivative of perturbation enstrophy. } 
The enstrophy equation is recalled below for the reader's convenience:
\begin{align}\label{eq:dtenstrophy}
\nonumber
\dtot{\Omega}{t}=&\frac{1}{Re}\mathfrak{R}\Bigl[\dddey\hat{\psi} \ddey\bar{\psi}\Bigr]^1_{-1}\\\nonumber&-\frac{1}{Re}\int^1_{-1} \big(3\alpha^2 \|\ddey\hat{\psi}\|^2+3\alpha^4\|\dey\hat{\psi}\|^2\\&+\alpha^6\|\hat{\psi}\|^2+\|\dddey\hat{\psi}\|^2\big)\ud y=\frac{1}{Re} H.
\end{align}
The procedure starts by writing $\dddey\hat\psi(\pm1,t)$ in terms of $\ddey\hat{\psi}$ and $\hat{\psi}$. In order to achieve this, the Orr-Sommerfeld equation (Eq. \ref{eq:OrrSomm2}) is multiplied by $e^{\epsilon \alpha y}$ and it is integrated over $[-1,1]$. By setting $\epsilon=1$, and $\epsilon=-1$, two independent equations are obtained. Then we solve for $\dddey\hat{\psi}(1,t)$ and $\dddey\hat{\psi}(-1,t)$. More in details, the Orr-Sommerfeld equation can be written as follows 
\begin{equation}\label{eq:OrrSomm2}
LL\hat{\psi}=Re\,\tilde{\sigma}L\hat{\psi}+i\alpha Re\,\tilde{M}\hat{\psi,}
\end{equation}
where
\begin{align}
L&=(\ddey-\alpha^2),\\
\tilde{\sigma}&=Re\Big(\dt+i\alpha U\Big),\\
\tilde{M}&=- U''.
\end{align}
Then, Eq. \ref{eq:OrrSomm2}() is multiplied by $e^{\epsilon \alpha y}\ud y$ and integrated over $[-1,1]$. The left hand side of Eq. (\ref{eq:OrrSomm2}), after integrating by parts and considering the boundary conditions
\begin{gather}\label{eq:OS_boundary_conditions}
 \hat\psi(\pm 1,t)=0,\\
 \dey\hat\psi(\pm 1,t)=0, 
 \end{gather}
 reads
\begin{align}
\nonumber&\int^1_{-1}  LL\hat{\psi} e^{\epsilon\alpha y}\ud y=\int^1_{-1} \ddddey \hat{\psi} e^{\epsilon\alpha y}\ud y -2\alpha^2 \int^1_{-1} \ddey\hat{\psi} e^{\epsilon\alpha y}\ud y \\ &+\alpha^4 \int^1_{-1}  \hat{\psi} e^{\epsilon\alpha y}\ud y = \left[(\dddey\hat{\psi}-\epsilon\alpha \ddey\hat{\psi})e^{\epsilon\alpha y}\right]_{-1}^1.
\end{align}
The right hand side of Eq. (\ref{eq:OrrSomm2}) requires some passages, since the operator $\tilde{\sigma}$ contains both a time derivative and the function $U(y)$:
\begin{flalign}
\nonumber&\int^1_{-1} \left(Re\,\tilde{\sigma}L\hat{\psi}+i\alpha Re\,\tilde{M}\hat{\psi}\right)e^{\epsilon\alpha y}\ud y\\=&\ Re\frac{\partial}{\partial t} \overbrace{\int^1_{-1} \ddey\hat{\psi} e^{\epsilon\alpha y}}^{\bf A}\ud y-Re\frac{\partial}{\partial t}\alpha^2 \int^1_{-1} \hat{\psi} e^{\epsilon\alpha y}\ud y\nonumber\\&+i\alpha Re \overbrace{\int^1_{-1} \ddey\hat{\psi} U e^{\epsilon\alpha y}}^{\bf B}\ud y-i\alpha^3 Re \int^1_{-1} \hat{\psi} U e^{\epsilon\alpha y}\ud y \nonumber\\ &-i\alpha Re \int^1_{-1} \hat{\psi}  U'' e^{\epsilon\alpha y}\ud y\nonumber\\ 
=&\ Re\frac{\partial}{\partial t}\alpha^2 \int^1_{-1} \hat{\psi} e^{\epsilon\alpha y}\ud y-Re\frac{\partial}{\partial t}\alpha^2 \int^1_{-1} \hat{\psi} e^{\epsilon\alpha y}\ud y\nonumber\\&+i\alpha Re\int^1_{-1} \hat{\psi} ( U''+2\epsilon\alpha U'+\alpha^2U)e^{\epsilon\alpha y}\ud y\nonumber\\&\nonumber-i\alpha^3 Re \int^1_{-1} \hat{\psi} U e^{\epsilon\alpha y}\ud y - i\alpha Re \int^1_{-1} \hat{\psi}  U'' e^{\epsilon\alpha y}\ud y\\=&\,2i\alpha^2\epsilon Re\int^1_{-1}  U'\,\hat{\psi} e^{\epsilon\alpha y}\ud y.
\end{flalign}
The terms $\bf A$ and $\bf B$ are evaluated separately by integrating by parts and using the boundary conditions:
\begin{align}
{\bf A}= &-\int^1_{-1} \dey \hat{\psi} \epsilon\alpha e^{\epsilon\alpha y}\ud y=\epsilon^2\alpha^2 \int^1_{-1} \hat{\psi} e^{\epsilon\alpha y}\ud y\nonumber\\=&\ \alpha^2 \int^1_{-1} \hat{\psi} e^{\epsilon\alpha y}\ud y, \\
{\bf B}= &-\int^1_{-1} \dey \hat{\psi} \dey (Ue^{\epsilon\alpha y})\ud y=\int^1_{-1} \hat{\psi} \ddey (Ue^{\epsilon\alpha y})\ud y\nonumber\\
= &\int^1_{-1} \hat{\psi} ( U''+2\epsilon\alpha U'+\epsilon^2\alpha^2U)e^{\epsilon\alpha y}\ud y.
\end{align}

The system of equations to find $\dddey\hat{\psi}(-1,t)$ and $\dddey\hat{\psi}(1,t)$ is 
\begin{align}\label{eq:vorticity_dynamical}
\begin{cases}
\left[(\dddey\hat{\psi}-\alpha \ddey \hat{\psi})e^{\alpha y}\right]_{-1}^1=2i\alpha^2 Re\int^1_{-1} U'\,\hat{\psi} e^{\alpha y}\ud y\\
\left[(\dddey\hat{\psi}+\alpha \ddey \hat{\psi})e^{-\alpha y}\right]_{-1}^1=-2i\alpha^2 Re\int^1_{-1} U'\,\hat{\psi} e^{-\alpha y}\ud y.
\end{cases}
\end{align}
Substituting these expressions in Eq. (\ref{eq:dtenstrophy}) and naming
\begin{align*}
a&=\cosh(2\alpha),\\
b&=\sinh(2\alpha),
\end{align*}
a new form for $H$ is obtained:
\begin{align}
\nonumber
\dtot{}{t}\Omega=&\frac{1}{Re} H=\frac{1}{Re} \Bigl[ \alpha a b^{-1}\Theta-\alpha b^{-1} \Phi+i\alpha^2 Re\,b^{-1}B+\\&-(I_3^2+3\alpha^2I_2^2+3\alpha^4I_1^2+\alpha^6I_0^2)\Bigr],\label{eq:modified_dtenstrophy}
\end{align}
where
\begin{align}
\label{eq:DueNove}
I_i^2=&\int^1_{-1} \partial_y^{(i)}\hat{\psi}  \partial_y^{(i)}\bar\psi \ud y,\\
\Theta= \,&\ddey\hat{\psi}(1,t)\ddey\bar{\psi}(1,t)+\ddey \hat{\psi}(-1,t)\ddey\bar{\psi}(-1,t) ,\\
\Phi= \,&\ddey\hat{\psi}(1,t)\ddey\bar{\psi}(-1,t)+ \ddey\hat{\psi}(-1,t)\ddey\bar{\psi}(1,t),\\
\label{eq:DueNoveB}B=& \int_{-1}^1 \Bigl[\hat{\psi}(y,t)\ddey\bar{\psi}(1,t)-\bar{\psi}(y,t)\ddey\hat{\psi}(1,t)\Bigr]U'\nonumber\\&\times\cosh[\alpha(y+1)]\ud y-\int_{-1}^1\Bigl[\hat{\psi}(y,t)\ddey\bar{\psi}(-1,t)\nonumber\\&
-\bar{{\psi}}(y,t)\ddey\hat{\psi}(-1,t)\Bigr]U'\cosh[\alpha(y-1)]\ud y.
\end{align}

$H$ depends on the parameters $\alpha$ and $Re$ and the function $\hat{\psi}$. In order to achieve conditions on $\alpha$ and $Re$ implying non-positivity of $H$, calculus of variations was used to maximize $H$ with respect to the function $\hat{\psi}$, \color{black} with vorticity values at the walls
\begin{gather}
p(t)=\ddey\hat{\psi}(1,t)\nonumber \\
q(t)=\ddey\hat{\psi}(-1,t)\nonumber
\end{gather}
being assigned.\\
\color{black}
Considering the part of $H$ depending on $\hat{\psi}$:
\begin{align}
H_{\hat{\psi}}=i\alpha^2Re\,b^{-1}B-(I_3^2+3\alpha^2I_2^2+3\alpha^4I_1^2+\alpha^6I_0^2),
\end{align}
introducing the variations on the perturbation, we evaluate
\begin{align*}
\underline{H} (\epsilon)=H_{\hat{\psi}}(\hat{\psi}+\epsilon\varphi) \quad \rightarrow \quad \frac{d\underline{H}}{d\epsilon}|_{\epsilon=0}=0.
\end{align*}
Calculus of variations leads to a sixth-order partial differential equation for the disturbance $\hat{\psi}$ which maximizes the enstrophy growth.  In the following, this particular function will be named $\hat{\psi}_m$
\begin{align}
\nonumber
&\partial_y^6  \hat{\psi}_m-3\alpha^2 \ddddey \hat{\psi}_m+3\alpha^4 \ddey  \hat{\psi}_m-\alpha^6 \hat{\psi}_m\\
=&\ i\alpha^2 Re\,b^{-1}U'(y)\Bigl\{p\cosh[\alpha(1+y)]-q\cosh[\alpha(1-y)]\Bigr\}.\label{eq:sixthorderpsi}
\end{align}

\begin{figure*}[h!]
	\includegraphics[width=0.83\textwidth]
	{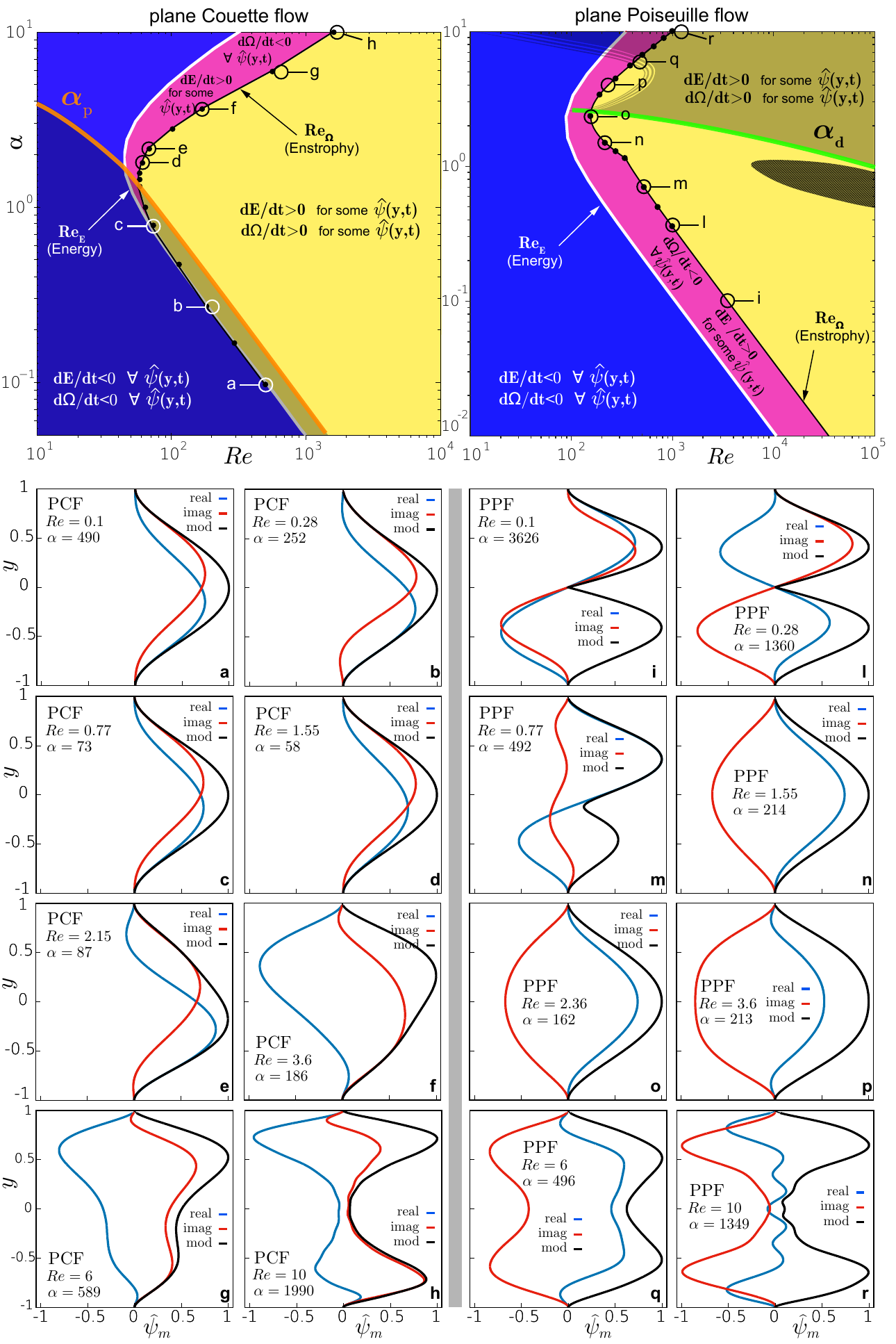}
	\caption{\color{black}\textbf{Enstrophy-rate optimal streamfunctions $\boldsymbol{\hat\psi_m}$ for $\boldsymbol{\alpha}, \mathbf{Re}$ points located close to the boundary of the monotonic decay region}. The points are labeled with lowercase letters from \textbf{a} to \textbf{r}. The figure reports optimal stream-functions computed with a numerical optimization procedure based on the initial-value problem \ref{eq:OrrSomm} (multi-parameter genetic optimization of 140 Chandrasekhar-Reid expansion coefficients). The points are located in the stability map at the margin of the region where the procedure found a positive enstrophy rate $(\Omega^{-1}\frac{\ud\Omega}{\ud t})_{max}$.  Please, remind that on the limit curve $(\Omega^{-1}\frac{\ud\Omega}{\ud t})_{max}=0$. Left side shows PCF (\textbf{a}-\textbf{h}), right side PPF (\textbf{i}-\textbf{r}). For details, the reader is remanded to the Appendix.}\label{fig:marginal_optimal}
\end{figure*}

\textcolor{black}{Notice that the solution $\hat \psi_m$ is the maximizing function for  assigned values of vorticity at the wall, $p$ and $q$. There will be specific values of $p$, $q$ yielding the absolute maximal enstrophy rate.  At this point, before proceeding with the solution, we highlight that
the optimization process of the enstrophy rate functional leading to  Eq. (\ref{eq:sixthorderpsi}) yields a basin of solutions $\hat\psi_m$ which is wider than that associated with the Orr-Sommerfeld equation. Essentially, this basin includes all stream functions which satisfy the physical boundary conditions (\ref{eq:OS_boundary_conditions}), among which the OS solutions represent a subset. This subset satisfies the \textit{dynamical conditions}  (\ref{eq:vorticity_dynamical}) \cite{synge1935}, which link the wall vorticity and its first derivative.
Such conditions are included in the maximized enstrophy functional in the new form (\ref{eq:modified_dtenstrophy}), but they are not imposed as a constraint in the calculus of variations leading to (\ref{eq:sixthorderpsi}). Imposing this constraint in the optimization procedure is mathematically nontrivial as it would require the solution of a 8\textsuperscript{th}-order PDE, with boundary conditions on $\hat \dddey\psi(\pm1,t)$ which are solution-dependent. However, not only Eq. (\ref{eq:sixthorderpsi}) is sufficient to find a lower bound for the enstrophy transient growth, but we also verified \textit{a posteriori} that the limit curve resulting from the solution of (\ref{eq:sixthorderpsi}) coincides with the best possible bound $Re_\Omega(\alpha)$, as defined by Eq. (\ref{eq:ReZ}). In order to prove this, a numerical procedure was set up. We used the solution of the OS initial-value problem obtained with our semi-analytical code  (published  in Appendix A of Ref. \cite{desanti2016}). By means of a genetic algorithm, the coefficients of the Chandrasekhar-Reid functions expansion are optimized, until the maximal enstrophy rate ($\Omega^{-1}\frac{\ud\Omega}{\ud t}$) at $t=0$ is found, for specified values of $Re$ and $\alpha$. Using the bisection method on $Re$, the limit of the enstrophy growth region was then found and compared with the results of the analytical procedure, see Fig.  \ref{fig:marginal_optimal}.  This allowed us to obtain the optimal functions $\hat\psi_m$ which also satisfy the dynamic condition  \ref{eq:vorticity_dynamical}. 
The numerical solution requires a large amount of computational resources (at least 140 Chandrasekhar-Reid modes need to be optimized). Enstrophy-rate optimal functions are shown in Fig. \ref{fig:marginal_optimal} below, and in Fig. SM 2.}

In the following, we proceed with the analytical computation of the monotonic decay region for the perturbation enstrophy.
The maximum value of $H$ corresponding to $\hat\psi_m$, named $H_{max}$, can be obtained by multiplying Eq. (\ref{eq:sixthorderpsi}) by $\overline{\psi}\,\ud y$, integrating over the range $(-1,1)$ and adding the complex conjugate. This yields
\begin{align}
&\Bigl[\ddey\hat{\psi}_m\dddey\bar\psi_m+\ddey\bar\psi_m\dddey\hat\psi_m\Bigr]^1_{-1}\nonumber\\&-2(I_3^2+3\alpha^2I_2^2+3\alpha^4I_1^2+\alpha^6I_0^2)\nonumber\\&=-i\alpha^2Re\,b^{-1}B,
\end{align}
and so from the definition of $H$:
\begin{align}
\label{eq:hmassimafinale}
\nonumber H_{max}&=\alpha a b^{-1} \Theta-\alpha b^{-1}\Phi+\frac{1}{2}i\alpha^2 Re\,b^{-1}B\\&-\frac{1}{2}\Bigl[\ddey\hat{\psi}_m\dddey\bar\psi_m+\ddey\bar\psi_m\dddey\hat\psi_m\Bigr]^1_{-1},
\end{align}
where here ${\hat\psi_m}$  is the stream-function which maximizes the enstrophy rate, solution of Eq. (\ref{eq:sixthorderpsi}).\\
The procedure followed up to this point leads to an expression for $H_{max}$ formally identical to that found by Synge (Eq. 2.12 in \cite{synge1938b}). The difference is that here, having adopted the  non-modal approach, $\hat{\psi}$ is time dependent. In the following, we solve the problem for PCF and then for PPF.


\begin{center}\small\textbf{A 1. Plane Couette flow: limit curve in the parameters space for the transient growth of traveling wave perturbations. Analytical method and final numerical optimization. Sufficient conditions for no enstrophy growth} 
\end{center}
In this section, conditions for no-growth of the perturbation enstrophy are derived for the plane Couette flow.  In this case $U(y)=y$, so that the equation \ref{eq:sixthorderpsi}, together with the boundary conditions, reads
\begin{align}\hspace{-2pt}
\label{eq:sixthorderpsiC}
\nonumber &\partial_y^6  \hat{\psi}_m(y,t)-3\alpha^2 \ddddey \hat{\psi}_m(y,t)+3\alpha^4 \ddey  \hat{\psi}_m(y,t)-\alpha^6 \hat{\psi}_m(y,t)\\
&=96i\,k\,\alpha^4\left\{p(t)\,\cosh[\alpha(1+y)]-q(t)\cosh[\alpha(1-y)]\right\},\\
&\hat{\psi}_m(\pm 1,t)=0,\\
&\dey \hat{\psi}_m(\pm 1,t)=0,\\
&\ddey \hat{\psi}_m(+1,t)=p(t),\\
&\ddey \hat{\psi}_m(-1,t)=q(t),
\end{align}
where  $k=Re/96 \alpha^2 b$.\\
We consider the homogeneous equation
\begin{align}
\label{eq:homequ}
\hat{\psi}_{m_H}^{(6)}-3\alpha^2\hat{\psi}_{m_H}^{(4)}+3\alpha^4\hat{\psi}_{m_H}^{(2)}-\alpha^6\hat{\psi}_{m_H}=0,
\end{align}
where $\hat{\psi}_{m_H}$ stands for the homogeneous solution.
Since the solutions of the characteristic equation are $+\alpha$ and $-\alpha$ both with multiplicity of 3, it is possible to write the solution as
\begin{align}
\hat{\psi}_{m_H}=(a_0+a_1\,y+a_2\,y^2) e^{-\alpha y}+(b_0+b_1\,y+b_2\,y^2)e^{\alpha y}.
\end{align}
Based on the form of the forcing term and in order to simpler compute of the constants when applying the boundary conditions, a different basis is chosen. In particular we write the solution as follows:
\begin{align}\label{eq:hom_integral}
\nonumber\hat{\psi}_{m_H}=&(a_0+a_1(1-y)+a_2(1-y)^2) \sinh[\alpha(1+y)]\\
&+(b_0+b_1(1+y)+b_2(1+y)^2)\sinh[\alpha(1-y)].
\end{align}
To show that this is indeed allowed, we proceed by proving that the two basis
\begin{align}
\mathscr{B}_1=(&e^{-\alpha y},e^{\alpha y},ye^{-\alpha y},ye^{\alpha y},y^2e^{-\alpha y},y^2e^{\alpha y})\\
\nonumber\mathscr{B}_2=(&\sinh[\alpha(1+y)],\sinh[\alpha(1-y)],(1-y)\sinh[\alpha(1+y)],\\\nonumber				&(1+y)\sinh[\alpha(1-y)],(1-y)^2\sinh[\alpha(1+y)],\\&(1+y)^2\sinh[\alpha(1-y)])
\end{align}
are linearly independent: to do so we write $\mathscr{B}_2=\mathscr{A}\mathscr{B}_1$, where
\begin{equation}
\mathscr{A}=\frac{1}{2}
\begin{pmatrix}
-e^{-\alpha} & e^\alpha & 0 & 0 & 0 & 0\\
e^\alpha & -e^{-\alpha} & 0 & 0 & 0 & 0\\
-e^{-\alpha} & e^\alpha & e^{-\alpha} & -e^\alpha & 0 & 0\\
e^\alpha & -e^{-\alpha} & e^\alpha & -e^{-\alpha} & 0 & 0\\
-e^{-\alpha} & e^\alpha & 2e^{-\alpha} & -2e^\alpha & -e^{-\alpha} & e^\alpha\\
e^\alpha & -e^{-\alpha} & 2e^\alpha & -2e^{-\alpha} & e^\alpha & -e^{-\alpha}
\end{pmatrix}.
\end{equation}
Since $\mathscr{A}$ is a triangular block matrix, the determinant is the product of the determinants of the three matrices on the diagonal. $Det(\mathscr{A})=4b^3\neq 0$, which implies linear independence. Thus, the general solution of Eq. (\ref{eq:homequ}) can also be  represented in the form \ref{eq:hom_integral}.\\
To solve equation \ref{eq:sixthorderpsiC}, a particular solution is to be found. We consider the forcing term as a superposition of two terms, one containing $\cosh[\alpha(1+y)]$ and the other one with $\cosh[\alpha(1-y)]$. We first find a particular solution  $\hat{\psi}_{m_{P1}}$ of the equation:
\begin{align}
\nonumber &\partial_y^6  \hat{\psi}_{m_{P1}}(y,t)-3\alpha^2 \ddddey \hat{\psi}_{m_{P1}}(y,t)+3\alpha^4 \ddey  \hat{\psi}_{m_{P1}}(y,t)\\&-\alpha^6 \hat{\psi}_{m_{P1}}(y,t)=96 i k \alpha^4 p\,\cosh[\alpha(1+y)].
\end{align}
We look for a solution with the form
\begin{align}
\hat{\psi}_{m_{P1}}(y,t)=-i\,k\,p\, y^3\left\{A_1 \sinh[\alpha(1+y)]+A_2 \cosh[\alpha(1+y)]\right\},
\end{align}
and obtain $A_2=0$ and $A_1=-2\alpha$, so the solution for this equation is:
\begin{align}
\hat{\psi}_{m_{P1}}(y,t)=2i\,k\,\alpha\,y^3 p\,\sinh[\alpha(1+y)].
\end{align}
Proceeding in the same way we find a second particular solution $\hat{\psi}_{m_{P2}}$ for:
\begin{align}
\nonumber &\partial_y^6  \hat{\psi}_{m_{P2}}(y,t)-3\alpha^2 \ddddey \hat{\psi}_{m_{P2}}(y,t)+3\alpha^4 \ddey  \hat{\psi}_{m_{P2}}(y,t)\\&-\alpha^6 \hat{\psi}_{m_{P2}}(y,t)=-96 i k \alpha^4 q\,\cosh[\alpha(1-y)],
\end{align}
leading to
\begin{align}
\hat{\psi}_{m_{P2}}(y,t)=2i\,k\,\alpha\,y^3 q\,\sinh[\alpha(1-y)].
\end{align}

The complete solution can be written as follows:
\begin{align}
\nonumber&\hat{\psi}_m(y,t)=\hat{\psi}_{m_{H}}(y,t)+\hat{\psi}_{m_{P1}}(y,t)+\hat{\psi}_{m_{P2}}(y,t)\\\nonumber
&=\left\{a_0+a_1(1-y)+a_2(1-y)^2+2ik\alpha y^3p\right\}\sinh[\alpha(1+y)]\\
&+\left\{b_0+b_1(1+y)+b_2(1+y)^2+2ik\alpha y^3 q\right\} \sinh[\alpha(1-y)].
\end{align}
By applying the boundary conditions it is possible to find the six constants (this was done by means of symbolic calculus via the Mathematica software):

\begin{flalign}\label{eq:ab_coefficients}
a_0=& -2 i \alpha k p, \nonumber\\
b_0=&\ 2 i \alpha k q, \nonumber\\
a_1=& -\{2 i \alpha (-48 \alpha^4 k p \cosh ^2(2 \alpha) +i \sinh ^2(2 \alpha)  (q+4 i \alpha^ 2 k p)\nonumber\\&
-4 \alpha^ 2 (4 \alpha^ 2 k p-2 \alpha k p \sinh (4 \alpha) +i q)+\cosh (2 \alpha) (64 \alpha^ 4 k q\nonumber\\&-4 i \alpha^ 2 p)-8 \alpha k q \sinh ^3(2 \alpha) +3 k p \sinh ^4(2 \alpha) +2 i \alpha p\nonumber\\&\times \sinh (2 \alpha) )\}/\{8 \alpha^ 4 \cosh (4 \alpha) -8 \alpha^ 3 (\alpha+\sinh (4 \alpha) )\nonumber\\&+12 \alpha^ 2 \sinh ^2(2 \alpha) -\sinh ^4(2 \alpha) \},\nonumber\\
a_2=&\{-64 i \alpha^ 5 k p \cosh ^2(2 \alpha) +2 \alpha^ 2 \sinh (4 \alpha)  (-16 i \alpha^ 2 k p+8 i \alpha \nonumber\\&\times k q \sinh (2 \alpha) +q )+8 \alpha^ 3 (q-8 i \alpha^ 2 k p)+8 \alpha^ 3 \cosh (2 \alpha)\nonumber\\&\times  (p+16 i \alpha^ 2 k q)+i \sinh (2 \alpha)  (64 \alpha^ 4 k q+\sinh (2 \alpha)  (16 \alpha^ 3 k p\nonumber\\&+\sinh (2 \alpha) (-48 \alpha^ 2 k q+ 12 \alpha k p \sinh (2 \alpha) +i p)+4 i \alpha q))\}\nonumber\\&/ \{32 \alpha^ 4 \cosh ^2(2 \alpha) -2 (8 \alpha^ 3 (2 \alpha+\sinh (4 \alpha) )-12 \alpha^ 2 \sinh ^2(2 \alpha)\nonumber\\&+\sinh ^4(2 \alpha) )\},\nonumber\\
b_1=&-\{2 i \alpha (16 \alpha^ 4 k q+48 \alpha^ 4 k q \cosh ^2(2 \alpha) -8 \alpha^ 3 k q \sinh (4 \alpha)\nonumber\\&+\sinh (2 \alpha)  (\sinh (2 \alpha)  (4 \alpha^ 2 k q +k \sinh (2 \alpha)  (8 \alpha p-3 q\nonumber\\&\times\sinh (2 \alpha) )+i p)+2 i \alpha q)-4 \alpha^ 2 \cosh (2 \alpha)  (16 \alpha^ 2 k p+i q)\nonumber\\&-4 i \alpha^ 2 p)\} / \{8 \alpha^ 4 \cosh (4 \alpha) -8 \alpha^ 3 (c+\sinh (4 \alpha) )+12 \alpha^ 2\nonumber\\&\times \sinh ^2(2 \alpha) -\sinh ^4(2 \alpha) \},\nonumber\\
b_2=&\{64 i \alpha^ 5 k q \cosh ^2(2 \alpha) +2 \alpha^ 2 \sinh (4 \alpha)  (16 i \alpha^ 2 k q-8 i \alpha k p\nonumber\\&\times \sinh (2 \alpha) +p)+8 \alpha^ 3 (p+8 i \alpha^ 2 k q)+8 \alpha^ 3 \cosh (2 \alpha)  (q-\nonumber\\&16 i \alpha^ 2 k p)-i \sinh (2 \alpha)  (64 \alpha^ 4 k p+\sinh (2 \alpha)  (16 \alpha^ 3 k q\nonumber\\&+\sinh (2 \alpha)  (-48 \alpha^ 2 k p+12 \alpha k q \sinh (2 \alpha) -i q)-4 i \alpha p))\}\nonumber\\& /\{32 \alpha^ 4 \cosh ^2(2 \alpha) -2 (8 \alpha^ 3 (2 \alpha+\sinh (4 \alpha) )-12 \alpha^2 \sinh ^2(2 \alpha)\nonumber\\& +\sinh ^4(2 \alpha) )\}.
\end{flalign}
Once the solution of Eq. (\ref{eq:sixthorderpsiC}) is available, it is possible to evaluate the maximal enstrophy growth, Eq. (\ref{eq:hmassimafinale}). We first evaluate $B$ in Eq. (\ref{eq:DueNoveB}) and the boundary term $\Bigl[\ddey\hat{\psi}_m\dddey\bar\psi_m+\ddey\bar\psi_m\dddey\hat\psi_m\Bigr]^1_{-1}$:
\begin{flalign}
&B= {}-4\,i\,k\left( Q_1 \Theta - Q_2 \Phi - \frac{Q_3}{k}\mathfrak{I}[p\overline{q}] \right),\\
&\Bigl[\ddey\hat{\psi}_m\dddey\bar\psi_m+\ddey\bar\psi_m\dddey\hat\psi_m\Bigr]^1_{-1}\nonumber\\&=\, 2 \left(F_1 \Theta - F_2 \Phi - kF_3\,\mathfrak{I}[p\overline{q}]\right),
\end{flalign}
where $\mathfrak{I}$ is the imaginary part and:
\begin{flalign}
\gamma_1=&384 \alpha ^3 (8 \alpha ^4 \cosh(4 \alpha ) + 12 \alpha ^2 \sinh(2 \alpha )^2 - \sinh(2 \alpha )^4\nonumber\\& - 8 \alpha ^3 (\alpha + \sinh(4 \alpha ))),\nonumber\\
2 \gamma_1 Q_1=&  -(8 \alpha  (3 + 16 \alpha ^2) (-9 + 24 \alpha ^2 + 64 \alpha ^4) -  32 \alpha  (-9 \nonumber\\& + 36 \alpha ^2 + 240 \alpha ^4 + 256 \alpha ^6) \cosh(4 \alpha ) + 24 \alpha  (-3 \nonumber\\& + 72 \alpha ^2 + 128 \alpha ^4) \cosh( 8 \alpha ) + (45 + 32 \alpha ^2 (27 \nonumber\\&+ 108 \alpha ^2 + 640 \alpha ^4  + 256 \alpha ^6)) \sinh(4 \alpha ) - 4 (9\nonumber\\& + 4 \alpha ^2 (27 + 252 \alpha ^2 + 64 \alpha ^4))\sinh(8 \alpha )\nonumber\\& +  9 \sinh(12 \alpha )),\nonumber\\
2 \gamma_1 Q_2=& -8 \alpha  (-9 - 432 \alpha ^2 - 288 \alpha ^4 + 512 \alpha ^6) \cosh(2 \alpha ) \nonumber\\&+4 \alpha  (-27 - 936 \alpha ^2 - 576 \alpha ^4 + 1024 \alpha ^6) \cosh(6 \alpha ) + \nonumber\\&  36 (\alpha + 8 \alpha ^3) \cosh(10 \alpha ) -  4 (45 + 4 \alpha ^2 (297 + 780 \alpha ^2 \nonumber\\& + 1536 \alpha ^4 + 640 \alpha ^6)) \sinh(2 \alpha ) -  2 (-45 + 4 \alpha ^2 (-243 \nonumber\\& + 4 \alpha ^2 (-231 + 64 \alpha ^2 (2 + \alpha ^2)))) \sinh(6 \alpha )  - 6 (3\nonumber\\& + 36 \alpha ^2  + 16 \alpha ^4) \sinh(10 \alpha )),\nonumber\\
\gamma_1 Q_3=&-((-12 \alpha  (9 + 56 \alpha ^2 + 64 \alpha ^4) - 16 \alpha  (-9 - 44 \alpha ^2 \nonumber\\& - 48 \alpha ^4) \cosh(4 \alpha )  +  4 \alpha  (-9 - 8 \alpha ^2) \cosh(8 \alpha ) \nonumber\\& -  16 \alpha ^2 (9 + 72 \alpha ^2 + 32 \alpha ^4) \sinh(4 \alpha ) \nonumber\\& + 72 \alpha ^2 \sinh(8 \alpha )),\nonumber\\
\gamma_2=&(-4 \sinh(2 \alpha )^3 + 16 \alpha ^2 (-4 \alpha  \cosh(2 \alpha ) \nonumber\\& + (3 + 4 \alpha ^2) \sinh(2 \alpha ))),\nonumber\\
\gamma_2 F_1=&\alpha ((3 - 48 \alpha ^2 + 64 \alpha ^4) \cosh(2 \alpha ) - 3 \cosh(6 \alpha )  \nonumber\\&+ 16 \alpha  (3 + 4 \alpha ^2) \sinh(2 \alpha )),\nonumber\\
\gamma_2 F_2=&(2 \alpha  (-3 + 16 \alpha ^2 + 32 \alpha ^4 + (3 + 8 \alpha ^2) \cosh(4 \alpha ) \nonumber\\& - 12 \alpha  \sinh(4 \alpha )),\nonumber\\
\gamma_2 F_3=&-((16 \alpha ^2 (36 \sinh(2 \alpha )^3 + 2 \alpha  ((9 + 144 \alpha ^2 + 64 \alpha ^4) \nonumber\\&\times \cosh(2 \alpha ) - 9 \cosh(6 \alpha ) + 4 \alpha  (-3 (7 \nonumber\\& + 8 \alpha ^2) \sinh(2 \alpha ) + \sinh(6 \alpha ))))).
\end{flalign}
\normalsize
This allows to rewrite Eq. (\ref{eq:hmassimafinale}) as:
\begin{align}\label{eq:Hmax_PCF}
\nonumber H_{max}=& \Phi \left\{ F_2-\frac{\alpha}{b}-\frac{Re^2 Q_2}{48b^2} \right\} - \Theta \left\{ F_1-\frac{\alpha\,a}{b}-\frac{Re^2 Q_1}{48b^2} \right\} \\
&\nonumber+ \mathfrak{I}[p\overline{q}] \left\{ k\,F_3-\frac{Re^2}{48b^2}\frac{1}{k}Q_3\right\}\\
=&\nonumber\Phi \left\{ F_2-\frac{\alpha}{b}-\frac{Re^2 Q_2}{48b^2} \right\} - \Theta \left\{ F_1-\frac{\alpha\,a}{b}-\frac{Re^2 Q_1}{48b^2} \right\} \\
&+ \mathfrak{I}[p\overline{q}] \left\{ \frac{Re}{96\alpha^2 b}\,F_3-\frac{2\alpha^2 Re}{b}Q_3\right\}.
\end{align}
The conditions for no-growth are obtained by looking for the region in the wavenumber-Reynolds  space where $H_{max}\leq0$. In the case of the plane Couette flow, this is done via a numerical optimization procedure as described below. Results are shown in Fig. \ref{fig:maps_summary}.

We proceed as follows. Supposing the existence of a limit curve $Re_\Omega(\alpha)$ which separates the region where $H_{max}>0$ from the region where $H_{max}<0$, we fix the wavenumber $\alpha$ and seek the Reynolds number at which $H_{max}=0$ for all the possible boundary terms $p$ and $q$. This is done through a genetic optimization algorithm implemented in Fortran 90, based on the open source software PIKAIA \cite{charbonneau1995,charbonneau1995user}. A wide range is set for the parameters $p$ and $q$. The functional (fitness function) to be minimized is $|H_{max}|$. We checked that increasing the numerical range for $p$ and $q$ did not influence the result.
A set of few wavenumbers was chosen and the computation was performed by optimizing over $p$, $q$ and $Re$, for each (fixed) wavenumber within the set. The Reynolds number giving the minimum $|H_{max}|$ from this procedure is represented with a black bullet, see Fig. \ref{fig:maps_summary}. 

\begin{center}\small\textbf{A 2. Plane Poiseuille flow: limit curve in the parameters space for the transient growth of traveling wave perturbations. Analytical method and final numerical optimization. Sufficient conditions for no enstrophy growth}
\end{center} 

In this section conditions for no enstrophy growth are found for the plane Poiseuille flow. We remind that the following analytical procedure for PPF is also present in Synge \cite{synge1938b} and it is here adopted to the non-modal formulation.  Let's consider  Eq. (\ref{eq:sixthorderpsi}) with $U'=-2y$. The solution to the homogeneous equation is the same as that for the Couette flow
\begin{align*}
\hat{\psi}_{m_H}=&\left[a_0+a_1(1-y)+a_2(1-y)^2\right] \sinh[\alpha(1+y)]\\
&+\left[b_0+b_1(1+y)+b_2(1+y)^2\right] \sinh[\alpha(1-y)].
\end{align*}
The forcing term differs from the PCF case  due to the presence of $U'$ in the right hand side. We seek a particular solution in the following form:
\begin{align}
\hat{\psi}_{m_{P}}=&-ik\left\{p\left[ \alpha y^4 \sinh[\alpha(1+y)]-6y^3\cosh[\alpha(1+y)]\right]\right.\nonumber\\
&\left.+q\left[\alpha y^4\sinh[\alpha(1-y)]+6y^3\cosh[\alpha(1-y)]\right]\right\}.
\end{align}
The complex constants $a_i$ and $b_i$ are determined by imposing the boundary conditions: vanishing $\hat{\psi}$ and $\dey\hat{\psi}$ and assigned values of $\ddey\hat{\psi}$. Direct calculation yields the following values:
\begin{align}
a_0=& ik\left( S_0p+T_0q\right), \nonumber\\
b_0=& ik\left( T_0p+S_0q\right), \nonumber\\
a_1=& \left(W_1+ikS_1\right)p+\left(V_1+ikT_1\right)q,\nonumber\\
b_1=& \left(V_1+ikT_1\right)p+\left(W_1+ikS_1\right)q,\nonumber\\
a_2=& \left(W_2+ikS_2\right)p+\left(V_2+ikT_2\right)q,\nonumber\\
b_2=& \left(V_2+ikT_2\right)p+\left(W_2+ikS_2\right)q.
\end{align}
The following real functions of $\alpha$ are involved:
\begin{align}
\label{eq:TreQuattro}
\gamma=& \tfrac{1}{4} \alpha^{-2} b^4-3b^2+4\alpha a b -4 \alpha^2b^2, \nonumber\\
W_1=& \gamma^{-1} (-b+2\alpha a), \quad V_1= \gamma^{-1} (-\tfrac{1}{2} \alpha^{-1} b^2 +2\alpha),\nonumber\\
W_2=& \gamma^{-1} (\tfrac{1}{8} \alpha^{-2}b^3-\alpha a),\nonumber\\
V_2=& \gamma^{-1} (\tfrac{1}{2} \alpha^{-1} b^2-\tfrac{1}{2}ab-\alpha ),\nonumber\\
S_0=& -6ab^{-1}+\alpha, \quad T_0= 6b^{-1},\nonumber\\
S_1=& \gamma^{-1} \left[ \tfrac{9}{2}\alpha^{-2}ab^3-\alpha^{-1}b^2(12+b^2)-6ab\right.\nonumber\\
&+\left.4c(6+7b^2)-72\alpha^2 ab +16\alpha^3 b^2 \right],\nonumber \\
T_1=& \gamma^{-1} \left[- \tfrac{9}{2}\alpha^{-2}b^3+12\alpha^{-1}ab^2+6b-4b^3-24\alpha a-8\alpha^2 b\right], \nonumber\\
S_2=&  -\tfrac{9}{2}\alpha^{-1} +3ab^{-1}-\tfrac{1}{2}S_1+\tfrac{1}{4}\alpha^{-1} b T_1,\nonumber\\
T_2=& \tfrac{9}{2} \alpha a-3b^{-1}-b-\tfrac{1}{4}\alpha^{-1}bS_1+\tfrac{1}{2}T_1,
\end{align}
where $a=\cosh(2\alpha)$ and $b=\sinh(2\alpha)$.
\\Once the maximizing perturbation $\hat{\psi}_m$ is available, it is possible to evaluate the maximal enstrophy growth (expression \ref{eq:hmassimafinale}). As done for PCF, we need to evaluate $B$ and $\Bigl[\ddey\hat{\psi}_m\dddey\bar\psi_m+\ddey\bar\psi_m\dddey\hat\psi_m\Bigr]^1_{-1}$. Given $\hat{\psi}_m$,
\begin{align}
&\int^1_{-1} \hat{\psi}_m y \cosh[\alpha(1+y)]\ud y=(P_1+ikQ_1)p-(P_2+ikQ_2)q,\nonumber\\
&\int^1_{-1} \hat{\psi}_m y \cosh[\alpha(1-y)]\ud y=(P_2+ikQ_2)p-(P_1+ikQ_1)q,
\end{align}
where $P_1$, $P_2$ are real constants and $Q_1$, $Q_2$, are given in terms of the constants just reported by
\begin{align}
\label{eq:TreSei}
Q_1=& \tfrac{1}{2}L_1\left\{ a\left(S_0+S_1+S_2\right)-\left(T_0+T_1+T_2\right) \right\} \nonumber\\&-\tfrac{1}{2} bL_2 \left(S_1+2S_2\right)+\tfrac{1}{2} L_3 \left(a S_2-T_2\right)+3aL_4\nonumber\\
&-\tfrac{1}{2}\alpha a L_5+\tfrac{6}{5}a+\tfrac{1}{3}b\left(T_1+2T_2\right),\nonumber\\
Q_2=& \tfrac{1}{2}L_1\left\{ S_0+S_1+S_2-a\left(T_0+T_1+T_2\right) \right\} \nonumber\\
&+\tfrac{1}{2} bL_2 \left(T_1+2T_2\right)+\tfrac{1}{2} L_3 \left(S_2-aT_2\right)+3L_4\nonumber\\
&-\tfrac{1}{2}\alpha L_5+\tfrac{6}{5}a-\tfrac{1}{3}b\left(S_1+2S_2\right),
\end{align}
where
\begin{align*}
L_n=\int^1_{-1} y^n e^{2\alpha y} \ud y, \quad n=1,2,\dots.
\end{align*}
Then, by direct calculation
\begin{flalign}
\label{eq:finaleEPois}
&B=-4ik\left( Q_1 \Theta -Q_2 \Phi\right),\\
\label{eq:finaleTreNovePois}
&\Bigl[\ddey\hat{\psi}_m\dddey\bar\psi_m+\ddey\bar\psi_m\dddey\hat\psi_m\Bigr]^1_{-1}=2\left(F_1 \Theta -F_2 \Phi\right),
\end{flalign}
where 
\begin{align}
\label{eq:TreDieci}
F_1=&\gamma^{-1}\Big( \frac{3}{4}a b^3 \alpha^{-1}-3b^2+3\alpha a b-4\alpha^2b^2-4\alpha^3 a b \Big),\\
F_2=&\gamma^{-1}\Big(  \frac{3}{4}b^3\alpha^{-1}-3a b^2+\alpha b (3+2b^2)+4\alpha^3b \Big).
\end{align}
Substituting Eqs. (\ref{eq:finaleEPois}) and (\ref{eq:finaleTreNovePois}) in Eq. (\ref{eq:hmassimafinale}), we eventually obtain
\begin{equation}
\label{eq:Hmax_PPF}
H_{max}=-\Theta(t)\Big\{F_1-\frac{\alpha a}{b}-\frac{Re^2 Q_1}{48b^2} \Big\}+\Phi(t)\Big\{F_2-\frac{\alpha}{b}-\frac{Re^2 Q_2}{48b^2} \Big\},
\end{equation}
which is an explicit function of $Re$, of the wavenumber $\alpha$ through $a,b,F_1,F_2,Q_1,Q_2$, and of the boundary terms $p=\ddey\hat{\psi}_m(1,t),m=\ddey\hat{\psi}_m(-1,t)$ through $\Theta$ and $\Phi$. Notice that the expression (\ref{eq:Hmax_PPF}) for the Poiseuille flow is apparently less complicated than the analogous found for the Couette flow in Eq. (\ref{eq:Hmax_PCF}). 
This simplification allows  to solve the problem for PPF in an analytical  way, as shown below.\par
From the definitions (\ref{eq:DueNove}) one can notice that $\Theta\ge 0$, $\Phi^2 \le \Theta^2$ for all times $t$ and for any complex value of $p$ and $q$. We see from Eq. (\ref{eq:Hmax_PPF}) that we can have $H_{max}\le 0$ for disturbances of wavelength $\lambda=2\pi/\alpha$ if $Re$ satisfies the two conditions:
\begin{align}
\label{eq:QuattroUno}
\begin{drcases}
&\frac{Re^2 Q_1}{48b^2}\le F_1-\frac{\alpha a}{b},\\
&\left[ F_2 -\frac{\alpha}{b}-\frac{Re^2 Q_2}{48b^2}\right]^2 \le \left[ F_1 -\frac{\alpha a}{b}-\frac{Re^2 Q_1}{48b^2}\right]^2 \\
\end{drcases},
\end{align}
where all  constants have already been defined.\\
To discuss these inequalities, we have to know the sign of $\gamma$ as defined in Eq. (\ref{eq:TreQuattro}). By expanding in series, one can notice that all coefficients are positive and, therefore, $\gamma$ is positive.\\
Writing $\xi=2\alpha$ so that $a=\cosh\xi$, $b=\sinh\xi$, and substituting Eq. (\ref{eq:TreQuattro}) in Eq. (\ref{eq:TreSei}), we obtain:
\begin{align}
\label{eq:QuattroCinque}
Q_1=&\frac{1}{\gamma\xi^7}b\Big\{ a(A'_0+A'_2b^2+A'_4b^4)+\xi b(B'_0+B'_2 b^2+B'_4b^4) \Big\},\nonumber\\
A'_0=&-624\xi^4-80\xi^6-\tfrac{8}{5}\xi^8,\nonumber\\
A'_2=&-1260\xi^2-1296\xi^4-148\xi^6-\tfrac{4}{3}\xi^8,\nonumber\\
A'_4=&204+12\xi^2,\nonumber\\
B'_0=& 1668\xi^2+672\xi^4+\tfrac{252}{5}\xi^6+\tfrac{4}{5}\xi^8 ,\nonumber\\
B'_2=&  12+1628\xi^2+\tfrac{2856}{5}\xi^4+\tfrac{64}{3}\xi^6,\nonumber\\
B'_4=&  -96,
\end{align}
and
\begin{align}
\label{eq:QuattroSei}
Q_2=&\frac{1}{\gamma\xi^7}b\Big\{ A''_0+A''_2b^2+A''_4b^4+\xi ab(B''_0+B''_2b^2) \Big\},\nonumber\\
A''_0=& -624\xi^4-80\xi^6-\tfrac{8}{5}\xi^8 ,\nonumber\\
A''_2=&  -1260\xi^2-1404\xi^4-228\xi^6-\tfrac{124}{15}\xi^8,\nonumber\\
A''_4=&  204-312\xi^2-108\xi^4-\tfrac{4}{3}\xi^8,\nonumber\\
B''_0=&  1668\xi^2+674\xi^4+\tfrac{252}{5}\xi^6+\tfrac{4}{5}\xi^8,\nonumber\\
B''_2=&  12+248\xi^2+\tfrac{96}{5}\xi^4.
\end{align}
From Eq. (\ref{eq:TreDieci}):
\begin{align}
\gamma \left(F_1-\frac{\alpha a}{b} \right)=& \xi^{-1} a b (3\xi^2 +b^2)-\xi^2-b^2(3+2\xi^2) ,\nonumber\\
\xi \gamma \left(F_2-\frac{\alpha}{b} \right)=& b(3\xi^2+\xi^4)+b^3(1+\xi^2)-a\xi(\xi^2+3b^2).
\end{align}
To solve the inequalities \ref{eq:QuattroUno}, it is convenient to define a function $\chi(\xi,\eta)$ with $\eta=0,\pm 1$ by
\begin{equation}
\label{eq:QuattroDieci}
\chi(\xi,\eta)=\frac{F_1-\alpha ab^{-1}+\eta \left(F_2-\alpha b^{-1}\right)}{Q_1 b^{-2}+\eta Q_2 b^{-2}}.
\end{equation}
Then, noting that $Q_1>0,\; Q_1+Q_2>0,\; Q_1-Q_2>0$,  the first of Eqs. (\ref{eq:QuattroUno}) can be written as inequality:
\begin{equation}
\label{eq:QuattroUndici}
\frac{Re^2}{48} \le \chi(\xi,0).
\end{equation}
The second inequality of \ref{eq:QuattroUno} becomes
\begin{equation}
\label{eq:QuattroTredici}
\left\{ \frac{Re^2}{48} - \chi(\eta,1)  \right\}\left\{ \frac{Re^2}{48} - \chi(\eta,-1)  \right\} \ge 0.
\end{equation}
As a result,  $d\Omega/dt \le 0$ for disturbances with a wavelength corresponding to an assigned value of $\xi$ provided that the following two conditions are satisfied:
\begin{itemize}
	\item $\dfrac{Re^2}{48}\le\chi(\xi,0)$;
	\item $\dfrac{Re^2}{48}$ is not between $\chi(\xi,-1)$ and $\chi(\xi,1)$.
\end{itemize}
The three functions $Re=[48\chi(\xi,0)]^\frac{1}{2}$; $[48\chi(\xi,+1)]^\frac{1}{2}$; $[48\chi(\xi,-1)]^\frac{1}{2}$  correspond to the black curves in Fig. \ref{fig:maps_summary} (\textbf{b}).  The region where  perturbations can experience transient  enstrophy growth is the yellow region in the same figure. Equation (\ref{eq:Hmax_PPF}) was also solved numerically as described in the above section for PCF, see the dotted curve in Fig. \ref{fig:maps_summary}(\textbf{b}). The nice match with the results from the two analytical conditions above allowed us to validate the algorithm, which was then used to solve the problem \ref{eq:Hmax_PCF} for the plane Couette flow, where analytical inequalities are not available. 
\FloatBarrier

\bibliographystyle{apsrev}

\end{document}